\documentclass[reprint,onecolumn,superscriptaddress,aip,showpacs,jap,10pt]{revtex4-1}

\usepackage{lineno}
\modulolinenumbers[5]

\usepackage[letterpaper, margin=1in]{geometry}
\usepackage{amsmath,amssymb}
\usepackage{graphicx}
\usepackage{bm}
\usepackage{siunitx}

\expandafter\ifx\csname package@font\endcsname\relax\else
 \expandafter\expandafter
 \expandafter\usepackage
 \expandafter\expandafter
 \expandafter{\csname package@font\endcsname}%
\fi

\renewcommand{\vec}[1]{\bm{#1}}
\newcommand{\ee}{\mathrm{e}}
\newcommand{\ii}{\mathrm{i}}
\newcommand{\dd}{\mathrm{d}}
\newcommand{\inc}{\mathrm{in}}
\newcommand{\sca}{\mathrm{sc}}
\newcommand{\tra}{\mathrm{tr}}

\newcommand{\LPW}{{\mathrm{LPW}}}
\newcommand{\SPW}{{\mathrm{SPW}}}

\DeclareMathOperator{\re}{Re}
\DeclareMathOperator{\im}{Im}

\newcommand{\bigO}{\mathrm{O}}
\newcommand{\SSS}{\mathrm{S}}
\newcommand{\LL}{\mathrm{L}}
\newcommand{\rad}{\mathrm{rad}}

\begin{document}

\title{Extended optical theorem in isotropic solids and its application to the elastic radiation force}

\author{J.~P. Le\~ao-Neto}
\affiliation{Physical Acoustics Group,
Instituto de F\'isica,
Universidade Federal de Alagoas, 
Macei\'o, AL 57072-970, Brazil}
\author{J. H. Lopes}
\affiliation{Grupo de F\'isica da Mat\'eria Condensada, N\'ucleo de Ci\^encias Exatas,
Universidade Federal de Alagoas,
Arapiraca, AL 57309-005, Brazil}
\author{G. T. Silva}
\email{glauber@pq.cnpq.br}
\affiliation{Physical Acoustics Group,
Instituto de F\'isica,
Universidade Federal de Alagoas, 
Macei\'o, AL 57072-970, Brazil}

\date{Ver. 20}

\begin{abstract}
The optical theorem is an important tool for scattering analysis in acoustics, electromagnetism, and quantum
mechanics. 
We derive an extended version of the optical theorem for the scattering of elastic waves by a spherical inclusion embedded in a linear elastic solid using a vector spherical harmonics representation of the waves. 
The sphere can be a rigid, empty cavity, elastic, viscoelastic, or layered material. 
The theorem expresses the extinction cross-section, i.e. the time-averaged power extracted from the incoming beam per its intensity, regarding the partial-wave expansion coefficients of the incident and scattered waves. 
We establish the optical theorem for a longitudinal spherically focused beam scattered by a sphere.
Moreover, we use the optical theorem formalism to obtain the radiation force exerted on an inclusion by an incident plane wave
and focused beam. 
Considering an iron sphere embedded in an aluminum matrix, we compute the scattering and elastic radiation force efficiencies.
In addition, the elastic radiation force is obtained on a stainless steel sphere embedded in a tissue-like medium (soft solid).
Remarkably, we find a relative difference of up to $98\%$ between our findings and previous lossless liquid models.
Regarding some applications, the obtained results have a direct impact on ultrasound-based elastography techniques, ultrasonic nondestructive testing,
as well as implantable devices activated by ultrasound.
\end{abstract}


\maketitle

\linenumbers




\section{Introduction}
Mechanical, electromagnetic, and quantum-mechanical wave scattering share some remarkable universal features.
A striking common characteristic among these fields is the optical theorem.
The original idea behind it was to relate the optical index of refraction of a medium to what has been extinct in the scattering process~\cite{Newton1976}.
For a traveling plane wave, the optical theorem states that the extinction cross-section, i.e. the time-averaged power extracted from
the incident wave by scattering and absorption per incident intensity, is tantamount 
the forward scattering function.
The theorem was initially stated for electromagnetic waves~\cite{Mie1908}.
In quantum-mechanics, it was derived by Feenberg~\cite{Feenberg1932}.
The optical theorem was also established for a plane electromagnetic~\cite{VanDeHulst1949,DeHoop1959}, plane
sound wave in an ideal fluid~\cite{DeHoop1959}, and plane elastic waves in solids~\cite{Barratt1965,Gubernatis1977,Varatharajulu1977,Korneev1993,Korneev1996}.
A generalized form of the optical theorem was proposed 
in the electron diffraction theory
using reciprocity relations~\cite{Glauber1953} and
in the acoustic scattering by objects with inversion
symmetry~\cite{Dassios1980,Marston2001}.
Furthermore, the generalized theorem was obtained for 
waves in a stratified medium~\cite{Ratilal2001}, surface waves~\cite{Kriegsmann1985,Halliday2009},
and  Raman scattering by
fractal clusters~\cite{Markel1991}.

It has been noticed that the ordinary optical theorem established for plane waves 
has some limitations.
It cannot be applied to beams with some transverse amplitude roll-off such as Gaussian beams~\cite{Lock1995}.
An extension of the optical theorem for nonplane wave scattering by a radially symmetric potential in quantum mechanics 
was presented in Ref.~\cite{Gouesbet2009}.
In this case, both
incident and scattered eigenstates
are expanded in
spherical function bases, allowing
the extinction cross-section be expressed
in terms of the expansion coefficients.
Another extended optical theorem was derived
for the on-axis scattering of a non-diffracting acoustic beam (such as Bessel beams) propagating in an ideal fluid~\cite{Zhang2013}.
This result was subsequently generalized for a scalar beam with arbitrary wavefront~\cite{Mitri2014}.
This problem has outstanding similarity with
the inelastic scattering of
quantum beams by a radial symmetric potential~\cite{Gouesbet2007}.
Extended optical theorems using the cylindrical wave decomposition has also been established for both acoustic~\cite{Mitri2015,Mitri2016}
 and electromagnetic~\cite{Mitri2015a} waves.

As noted by Newton~\cite{Newton1976}, the optical theorem accounted for dispersion of light propagating in a material.
A description of x-rays dispersion was also provided based on similar ideas~\cite{DeL.Kronig1926}.
It also served as the foundation of the connection between dispersion relation and causality~\cite{Toll1956}.
A wide variety of applications of the optical theorem includes
 phase shift estimation from measurements of the differential scattering cross-section in quantum mechanics~\cite{Newton1968},
 evaluation of cracks in elastic solids~\cite{Kraut1976,Kitahara1997},
 diffraction tomography~\cite{Carney1999},
 analysis of attenuation effects from scatterers~\cite{Groenenboom1995},
 Green's function reconstruction
 in inhomogeneous elastic solid medium~\cite{Margerin2011}, seismic interferometry~\cite{Wapenaar2010}, and
 calculation of energy loss in solids with dislocation~\cite{Maurel2008},
 to name a few.
In ultrasonic
nondestructive testing (NDT)~\cite{Krautkramer1990},  
an ultrasound wave is employed to investigate solid structures with inclusions, dislocation, and microcracks~\cite{Smyshlyaev1994}.

Motivated by the wide range of applications that the optical theorem may bring to elastodynamics and NDT,
we developed the extended formalism applicable
to any longitudinal or shear ultrasound beam of an arbitrary wavefront.
Also, we apply the optical theorem framework to derive the mean force exerted on the inclusion in a solid matrix
by elastic waves.
In lossless fluids, this force, known as the acoustic radiation force, has been theoretically analyzed in Refs.~\cite{Hasegawa1969,Chen1996,Marston2006,Mitri2009,Silva2011a,Azarpeyvand2012,Sapozhnikov2013,Baresch2013}.
In solids,  we refer to it as \emph{the elastic radiation force}.
This force plays a key role in some elastography methods~\cite{Wells2011,Sarvazyan2011,Palmeri2011,Bercoff2004}.
It is also related to the ultrasound-activation mechanism for implanted devices~\cite{Ordeig2016}. 
Moreover, the displacement induced by the elastic radiation force on a particle embedded in a viscoelastic gel 
has been experimentally measured~\cite{Aglyamov2007,Andreev2016}.

With the developed formalism, we revisit the scattering of longitudinal and shear plane waves by a spherical inclusion.
We derive the optical theorem for the scattering of a longitudinal spherically focused beam
by an on-focus sphere.
We numerically compute the extinction and 
radiation force efficiencies for the scattering of the beams mentioned above.
The incident waves are scattered by an iron sphere embedded in an aluminum matrix.
The role of mode conversion in scattering is featured.
Additionally, we consider a stainless steel sphere in a tissue-like medium (soft solid).
The obtained radiation force considerably deviates from that computed based on the previous lossless liquid model for the medium~\cite{Chen1996}.
A $98\%$-relative difference is found between our findings and previous models.
Thus, estimating the radiation force in soft solids assuming a liquid medium may lead
to an enormous error.
This work was partially presented in the 5th Joint Meeting of the Acoustical Society of America and Acoustical Society of Japan,
Honolulu, Hawaii, 2016.

\section{Theory}

\subsection{Wave propagation and scattering}

Consider an unbounded medium composed of an isotropic elastic solid 
with density $\rho_0$. 
The displacement vector of a point at position vector $\vec{r}$ is denoted by $\bm{u}$. 
The stress induced by small perturbation in the medium is based on the Hooke's law,~\cite{Fung2001}
\begin{equation}
\label{stress-strain}
 \bm{\sigma} = K_0 \left(\nabla\cdot \vec{u}\right)\mathbf{I}
 + \mu_0 \left(\nabla \bm{u}
  +\nabla \bm{u}^T\right),
\end{equation}
where the constants
$K_0$ and $\mu_0$ are, respectively, the bulk and shear modulus,
$\mathbf{I}$ is the second-rank unit tensor,
$\nabla \bm{u}$ is a second-rank tensor, and
 the superscript $^T$ denotes the transpose operation.
The conservation of linear momentum requires 
\begin{equation}
\label{motion}
\rho_0 \partial_t^2 \bm{u} = \nabla \cdot 
\bm{\sigma}.
\end{equation}
Here, we are using the shorthand notation
$\partial_t=\partial/\partial t$.
Substituting Eq.~(\ref{stress-strain})
into this equation yields
\begin{equation}
\label{WE1}
\rho_0\partial_t^2\bm{u} =\left (K_0 + \mu_0\right)\nabla(\nabla \cdot \bm{u})+\mu_0\,\nabla^2\bm{u}.
\end{equation}
This is the wave equation supporting longitudinal (L) and shear (S) waves.
By employing the identity $\nabla^2\bm{u}=\nabla(\nabla \cdot \bm{u})- \nabla \times \nabla \times \bm{u}$,
we obtain 
\begin{equation}
\label{equationofmotion}
\partial_t^2 \bm{u} =c_\LL^2\nabla(\nabla \cdot \bm{u})-c_\SSS^2\nabla \times \nabla \times\bm{u},
\end{equation}
where the longitudinal and shear speed of sound are, respectively,
\begin{equation}
\label{sound-speeds}
 c_\LL = \sqrt{\frac{K_0 + 2\mu_0}{\rho_0}}, \quad
 c_\SSS = \sqrt{\frac{\mu_0}{\rho_0}}.
 \end{equation}
Note that the longitudinal speed of sound is  larger than its shear counterpart, $c_\LL>c_\SSS$. 
 
Assume that a time-harmonic displacement of angular frequency $\omega$ is induced  in the solid.  
Using the Helmholtz decomposition theorem~\cite{Morse1953}, 
we can express the 
displacement amplitude vector 
as
\begin{equation}
\label{Helm_eq}
\vec{u}= \left[\vec{u}^{(\LL)}+\vec{u}^{(\SSS)}\right] \ee^{-\ii \omega t},
\end{equation}
where 
`$\ii$' is the imaginary-unit,
$\vec{u}^{(\LL)}$ (longitudinal component) is irrotational $\nabla\, \times \,\vec{u}^{(\LL)} = 0$,
and
$\vec{u}^{(\SSS)}$ (shear component) 
is divergenceless $\nabla\, \cdot\, \vec{u}^{(\SSS)}=0$.

Hereafter, we consider the longitudinal $\vec{u}^{(\LL)}$ and shear $\vec{u}^{(\SSS)}$ displacements as normalized quantities to
the displacement magnitude $u_0$.
Inserting Eq.~(\ref{Helm_eq}) into Eq.~(\ref{equationofmotion}), we find the 
vector Helmholtz equations,
\begin{subequations}
\label{helm}
\begin{eqnarray}
\left(\nabla^2 + k_\LL^2\right) \vec{u}^{(\LL)} &=& 0, \\
\left(\nabla^2 + k_\SSS^2\right) \vec{u}^{(\SSS)} &=& 0,
\end{eqnarray} 
\end{subequations}
where $k_\LL=\omega/c_\LL$
and $k_\SSS=\omega/c_\SSS$ are
the longitudinal and shear wavenumbers, respectively.

\begin{figure}
\centering
\includegraphics[scale=.5]{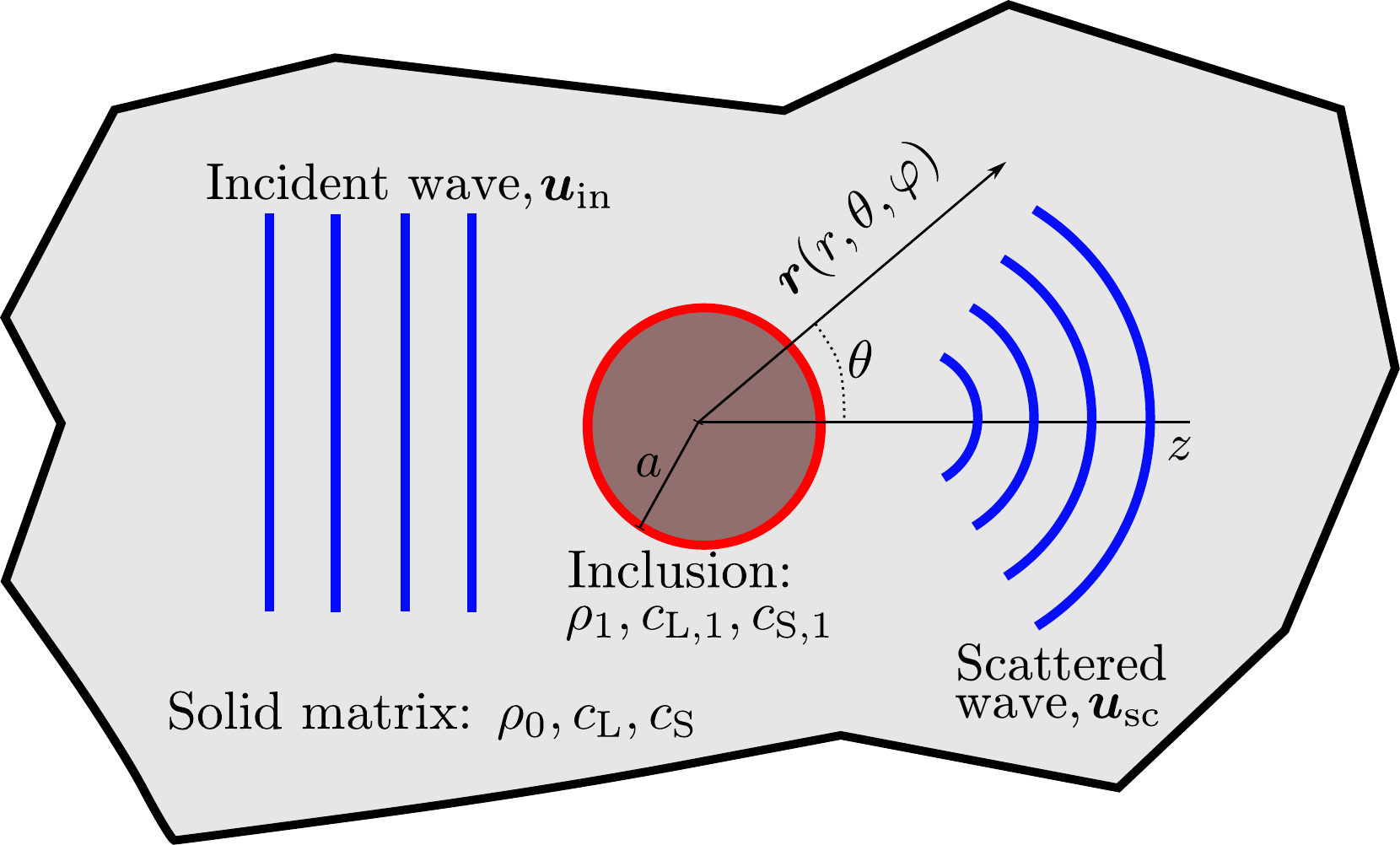} 
\caption{\label{fig:scattering}
(Color online) 
The sketch of the scattering problem.
An arbitrary incident wave denoted by blue vertical bars
is scattered by a spherical inclusion of radius $a$, density $\rho_1$, longitudinal $c_{\LL,1}$ and shear $c_{\SSS,1}$
speed of sound. The sphere is  embedded in a solid matrix of density $\rho_0$, longitudinal $c_{\LL}$ and shear $c_{\SSS}$
speed of sound.
The scattered waves are depicted by blue arches.
The axial $z$-direction and  position vector in spherical coordinates $(r, \theta, \varphi)$ are illustrated.} 
\end{figure}

In the presence of an inclusion,
both  incident longitudinal and shear waves will be scattered.
In Fig.~\ref{fig:scattering}
we illustrate the scattering problem for a sphere of radius $a$ and density $\rho_1$ embedded in an elastic solid.
The  longitudinal and shear speed of sound of the sphere are denoted by $c_{\LL,1}$ and $c_{\SSS,1}$,
respectively.
For convenience, we adopt spherical coordinates  
$(r,\theta, \varphi)$, where $r$ is the radial distance, $\theta$ and $\varphi$ are the polar and azimuthal angles.
The unit-vectors in spherical coordinates are denoted by $\vec{e}_r$, 
$\vec{e}_\theta$, and $\vec{e}_\varphi$.
In terms of the incident (in) and scattered (sc) fields, the vector displacements are expressed as
\begin{subequations}
\begin{eqnarray}
\vec{u}^{\mathrm{(L)}} &=& 
\vec{u}^{\mathrm{(L)}}_\mathrm{in} +
\vec{u}^{\mathrm{(L)}}_\mathrm{sc},\\
\vec{u}^{\mathrm{(S)}} &=& 
\vec{u}^{\mathrm{(S)}}_\mathrm{in} +
\vec{u}^{\mathrm{(S)}}_\mathrm{sc}.
\end{eqnarray}
\end{subequations}

Let us first discuss the solutions of the Helmholtz vector equations in~(\ref{helm}) for the incident wave.
They should be finite everywhere in space.
The regular base solution of the vector Helmholtz equations in spherical coordinates are 
given in terms of the Hansen vectors~\cite[p.~1799]{Morse1953a}
\begin{subequations}
\label{Hensen_vectors}
\begin{eqnarray}
\label{L1}
\vec{L}_{nm}^{(1)} &=& \nabla_\LL \left[j_n(k_\LL r) Y_{n}^m(\theta,\varphi)\right] =
j_n'(k_\LL r) \vec{Y}_{nm}(\theta,\varphi) +\frac{j_n(k_\LL r)}{k_\LL r} \vec{\Psi}_{nm}(\theta,\varphi), \\
\label{M1}
\vec{M}_{nm}^{(1)} &=& \nabla_\SSS \times \left[k_\SSS r \, j_n(k_\mathrm{S} r) \vec{Y}_{nm}(\theta,\varphi) \right]
=
-j_n(k_\SSS r)\vec{\Phi}_{nm}(\theta, \varphi),\\
\label{N1}
\vec{N}_{nm}^{(1)} &=& \nabla_\SSS \times \vec{M}_{nm}^{(1)}
=n(n+1)\frac{j_n(k_\SSS r)}{k_\SSS r}\vec{Y}_{nm}(\theta,\varphi) + \frac{\partial_r[r j_n(k_\SSS r)]}{k_\SSS r}\vec{\Psi}_{nm}(\theta,\varphi),
\end{eqnarray}
\end{subequations}
where $\nabla_j=k_j^{-1} \nabla$ with $j\in \{\LL,\SSS\}$, $j_n$ is the $n$th-order spherical Bessel function, the prime symbol means differentiation.
The vector spherical harmonics in these equations are defined as~\cite{Barrera1985}
\begin{equation}
\label{VSH}
\vec{Y}_{nm}(\theta,\varphi)\equiv Y_{n}^m(\theta,\varphi) \vec{e}_r,\quad
\vec{\Psi}_{nm}(\theta,\varphi) \equiv r \nabla Y_{n}^m(\theta,\varphi),\quad
\vec{\Phi}_{nm}(\theta,\varphi) \equiv \vec{r} \times \nabla Y_{n}^m(\theta,\varphi).
\end{equation}
The spherical harmonic of  $n$th-order and $m$th-degree
is 
\begin{equation}
Y_n^m(\theta,\varphi) =
\sqrt{\frac{(2n+1)}{4\pi} \frac{(n-m)!}{(n+m)!}}P_{n}^m(\cos \theta) \,\ee^{\ii m \varphi}, 
\label{SH}
\end{equation} 
where~\cite{Abramowitz1964}
\begin{equation}
P_{n}^m(z)=\frac{(-1)^m}{2^n n!} (1-z^2)^{m/2} \frac{\dd^{n+m}}{\dd z^{n+m}}(z^2-1)^n
\end{equation}
being  the associated Legendre polynomial of $n$th-order and $m$th-degree.

The partial-wave expansion of the longitudinal and shear displacements 
are expressed as
\begin{subequations}
\label{set_solution}
\begin{eqnarray}
\label{uinL}
&&\vec{u}_\inc^{(\LL)}=
\sum_{n,m} 
 a_{nm}^{(\LL)} \vec{L}_{nm}^{(1)}(k_\LL r, \theta, \varphi),
 \\
&&\vec{u}_\inc^{(\SSS)} = 
\sum_{n,m}
   a_{nm}^{(\SSS,1)}  \vec{M}_{nm}^{(1)}(k_\SSS r, \theta, \varphi)
+
a_{nm}^{(\SSS,2)}  \vec{N}_{nm}^{(1)}(k_\SSS r, \theta, \varphi), 
\label{uinMN}
\end{eqnarray}
\end{subequations}
where 
$\sum_{n,m}=\sum_{n=0}^\infty\sum_{m=-n}^n$,
 $a_{nm}^{(\LL)}$ is
the expansion (beam-shape) coefficient of the longitudinal wave,
while
 $a_{nm}^{(\SSS,1)}$
and $a_{nm}^{(\SSS,2)}$ are the expansion coefficients of the of the first- and second-type shear wave.
The beam-shape coefficients can be determined by
using the orthogonal relations in (\ref{app:ortho})
into Eqs.~(\ref{uinL}) and (\ref{uinMN}).
The result yields
\begin{subequations}
\begin{eqnarray}
a_{nm}^{(\LL)}&=& \frac{1}{j_n'(k_\LL r)}\oint_{4\pi}\vec{u}_\inc^{(\LL)}( k_\LL r, \theta, \varphi) \cdot
 \vec{Y}_{nm}^*( \theta, \varphi)
 \, \dd \Omega, 
  \label{anm_L_1}
  \\
  a_{nm}^{(\SSS,1)} &=& -\frac{1}{n(n+1) j_n(k_\SSS r)}
  \oint_{4 \pi}\vec{u}_\inc^{(\SSS)}( k_\SSS r, \theta, \varphi) \cdot
   \vec{\Phi}_{nm}^*(\theta, \varphi)
   \, \dd \Omega,
   \label{anm_S_2}
   \\
a_{nm}^{(\SSS,2)} &=& 
\frac{k_\SSS r}{n(n+1) j_n(k_\SSS r)}
  \oint_{4 \pi}\vec{u}_\inc^{(\SSS)}( k_\SSS r, \theta, \varphi) \cdot
   \vec{Y}_{nm}^*(\theta, \varphi)
   \, \dd \Omega,
    \label{anm_S_1}
\end{eqnarray}
\end{subequations}
where asterix denotes complex conjugation,
 the centered dot  means scalar product, and
 $\dd \Omega = \sin \theta \,\dd \theta \dd \varphi$ is the differential solid angle.
 We note that the beam-shape coefficients can also be computed by numerical schemes for a given longitudinal or shear incident
 vector displacements.
 This is particularly useful in off-axial scattering problems~\cite{Silva2011,Mitri2011,Silva2015}.

The asymptotic form of the incident displacement fields at the farfield ($k_\LL r, k_\SSS r \rightarrow \infty$)
is necessary to derive the extended optical theorem later.
Thus, using the asymptotic expression~\cite{Abramowitz1964}
$j_n(x)=
x^{-1}\cos\left[ x - (n+1) \pi/2 \right]+\bigO\left(x^{-2}\right)
$
into (\ref{set_solution}), we arrive at
\begin{subequations}
\begin{eqnarray}
\label{uinL_far}
\vec{u}_\inc^{(\LL)} &=& -\frac{1}{k_\LL r} 
\sum_{n,m} a_{nm}^{(\LL)}
  \sin \left[k_\LL r -\frac{(n+1) \pi}{2}\right] \vec{Y}_{nm}(\theta,\varphi),\\
\nonumber
\vec{u}_\inc^{(\SSS)} &=&
-\frac{1}{k_\SSS r}
\sum_{n,m}
 \biggl\{
   a_{nm}^{(\SSS,1)}  \cos \left[k_\SSS r -\frac{(n+1) \pi}{2}\right] \vec{\Phi}_{nm}(\theta,\varphi)
   \nonumber \\
&+& a_{nm}^{(\SSS,2)}   \sin \left[k_\SSS r -\frac{(n+1) \pi}{2}\right] \vec{\Psi}_{nm}(\theta,\varphi)
\biggr\}.
\label{uinMN_far}
\end{eqnarray}
\end{subequations}


We turn our attention to the longitudinal and shear scattered waves.
The displacement associated to these waves can also be expanded into partial-waves likewise 
Eqs.~(\ref{uinL}) and (\ref{uinMN}).
The domain of the scattered displacement fields
should
excludes the scatterer.
In this case, the radial component of the vector spherical functions should satisfy  
radiation conditions~\cite{Pao1976}.
Hence, the regular spherical Bessel functions $j_n$ in the basis vector functions given in~(\ref{set_solution})
should be replaced by the spherical Hankel functions of the first type $h_n^{(1)}$.
By so doing, we introduce the singular Hensen vectors
\begin{subequations}
\label{Hensen_vectors_sing}
\begin{eqnarray}
\label{L2}
\vec{L}_{nm}^{(2)} &=& 
{h_n^{(1)}}'(k_\LL r) \vec{Y}_{nm}(\theta,\varphi) +\frac{h_n^{(1)}(k_\LL r)}{k_\LL r} \vec{\Psi}_{nm}(\theta,\varphi), \\
\label{M2}
\vec{M}_{nm}^{(2)} &=& 
-h_n^{(1)}(k_\SSS r)\vec{\Phi}_{nm}(\theta, \varphi),\\
\label{N2}
\vec{N}_{nm}^{(2)} &=& n(n+1)\frac{h_n^{(1)}(k_\SSS r)}{k_\SSS r}\vec{Y}_{nm}(\theta,\varphi) + \frac{\partial_r[r h_n^{(1)}(k_\SSS r)]}{k_\SSS r}\vec{\Psi}_{nm}(\theta,\varphi).
\end{eqnarray}
\end{subequations}
Now, we express the scattering displacements as
\begin{subequations}
\label{uscattered}
\begin{eqnarray}
\label{uscL}
&&\vec{u}_\sca^{(\LL)}( k_\LL r, \theta, \varphi) =
 \sum_{n,m}
  s_{nm}^{(\LL)} \vec{L}_{nm}^{(2)}(k_\LL r, \theta, \varphi),
  \\
&&\vec{u}_\sca^{(\SSS)}(k_\SSS r, \theta, \varphi) =
\sum_{n,m}
s_{nm}^{(\mathrm{S},1)}  \vec{M}_{nm}^{(2)}(k_\SSS r, \theta, \varphi)
+
s_{nm}^{(\mathrm{S},2)}  \vec{N}_{nm}^{(2)}(k_\SSS r, \theta, \varphi),
\label{uscMN}
\end{eqnarray}
\end{subequations}
where $s_{nm}^{(\LL)}$ is the longitudinal, and $s_{nm}^{(\SSS,1)}$ and $s_{nm}^{(\SSS,2)}$
are the first- and second-type shear scattering coefficients.
These coefficients can be determined by applying the  continuity condition on the displacement and stress fields across the 
inclusion's boundary as will be shown later.

Using
the asymptotic form of the spherical Hankel function for large arguments~\cite{Abramowitz1964}
$h_n^{(1)}(x)=
(-\ii)^{n+1} \ee^{\ii x}/ x +\bigO\left(x^{-2}\right)
$ 
into Eqs.~(\ref{uscL}) and (\ref{uscMN}),
we obtain the asymptotic behavior of the scattered waves at the farfield,
\begin{equation}
\label{uscL_far}
\vec{u}_\sca^{(j)} =
\frac{\ee^{\ii k_j r}}{k_j r}
\vec{f}^{(j)}(\theta, \varphi)
+ \bigO\left[(k_j r)^{-2}\right]
, \quad
j \in \{\LL,\SSS\}.
\end{equation}
The scattering form functions are given by
\begin{subequations}
\begin{eqnarray}
\label{fL}
\vec{f}^{(\LL)}(\theta, \varphi) &=&
\sum_{n,m}  
\ii^{-n}
s_{nm}^{(\LL)}
 \vec{Y}_{nm}(\theta,\varphi),\\
\vec{f}^{(\SSS)}(\theta, \varphi) &=& 
\sum_{n,m}
\ii^{-n}  \bigl[
\ii s_{nm}^{(\SSS,1)} \vec{\Phi}_{nm}(\theta,\varphi)
+
s_{nm}^{(\SSS,2)} \vec{\Psi}_{nm}(\theta,\varphi)
\bigr].
\label{fS}
\end{eqnarray}
\end{subequations}
Using the equations in~(\ref{VSH_forward}) and (\ref{VSH_backward}), we derive scattering function in the forward and backward directions as
\begin{subequations}
\begin{eqnarray}
\label{fL_forward}
\vec{f}^{(\LL)}(\theta=0,\pi,\varphi=0) 
&=& 
\sum_{n=0}^\infty  
\epsilon_n \ii^{n}
\sqrt{\frac{2n +1}{4 \pi}} 
s_{n,0}^{(\LL)}\,
 \vec{e}_z,\\
 \nonumber
 \vec{f}^{(\SSS)}(\theta=0,\pi,\varphi=0) &=& -
\sum_{n=1}^\infty 
\frac{\epsilon_n \ii^{n}}{2}
\sqrt{\frac{(2n +1)n(n+1)}{4 \pi }}  
\left[
\biggl(
s_{n,-1}^{(\SSS,1)} 
+ s_{n,1}^{(\SSS,1)}
- s_{n,-1}^{(\SSS,2)}
+ s_{n,1}^{(\SSS,2)}\right)
 \vec{e}_x \\
&\mbox{}& -\ii
 \left(
 s_{n,-1}^{(\SSS,1)}
 -s_{n,1}^{(\SSS,1)}
 - s_{n,-1}^{(\SSS,2)}
 -  s_{n,1}^{(\SSS,2)}\right)
  \vec{e}_y\biggr],
  \label{fS_forward}
\end{eqnarray}
\end{subequations}
where
$\epsilon_n =-1, (-1)^n$ if $\theta=0, \pi$;
and  $\vec{e}_x$, $\vec{e}_y$, and $\vec{e}_z$
are the Cartesian unit-vectors.

Both longitudinal and shear waves might be transmitted into the inclusion.
Since the transmission waves should be regular 
everywhere inside the inclusion, we have
\begin{subequations}
\label{u_transmitted}
\begin{eqnarray}
\label{utL}
&&\vec{u}_\tra^{(\LL)}( k_\LL r, \theta, \varphi) =
\sum_{n,m} t_{nm}^{(\LL)} \vec{L}_{nm}^{(1)}(k_\LL r, \theta, \varphi),
\\
&&
\vec{u}_\tra^{(\SSS)}(k_\SSS r, \theta, \varphi) =
 \sum_{n,m}
\left[
t_{nm}^{(\SSS,1)}  \vec{M}_{nm}^{(1)}(k_\SSS r, \theta, \varphi)
+
t_{nm}^{(\SSS,2)}  \vec{N}_{nm}^{(1)}(k_\SSS r, \theta, \varphi)
\right],
\label{utMN}
\end{eqnarray}
\end{subequations}
where $t_{nm}^{(\LL)}$ is the longitudinal, and $t_{nm}^{(\SSS,1)}$ and $t_{nm}^{(\SSS,2)}$ are the shear transmission coefficients.
They can also be determined by applying the continuity condition of stresses and displacements across the inclusion's surface.

\subsection{Extended optical theorem}
\label{sec:opt_theorem}

Mechanical waves carry energy while propagating.
When a wave encounters an inclusion, part of its energy is extincted due to scattering and absorption within the inclusion. 
To analyze this phenomenon 
it is useful to define the absorption $\sigma_\mathrm{abs}$ and
scattering
$\sigma_\mathrm{sca}$ cross-section areas as
\begin{equation}
\label{sigmas}
\sigma_{\mathrm{abs},\mathrm{sca}} \equiv \frac{P_{\mathrm{abs},\mathrm{sca}}}{I_0},
\end{equation}
where $P_\mathrm{abs}$
and
$P_\mathrm{sca}$
are the time-averaged absorbed and scattering power, and
$I_0$ is time-averaged characteristic intensity of the incident beam.
This means that
the total absorption (scattering) power is equal to the incident intensity $I_0$ projected onto the absorption 
(scattering) cross-section area.
From the conservation of energy principle, 
the power removed (extinct) from the incident wave is
$P_\textrm{ext}= P_\textrm{abs} + P_\textrm{sca}$.
Therefore,
the  extinction cross-section is given by
\begin{equation}
\label{sigma_ext}
\sigma_\mathrm{ext}=\sigma_\mathrm{abs} + \sigma_\mathrm{sca}.
\end{equation}
It is useful to introduce the absorption, scattering, and extinction efficiencies as 
their respectively cross-sections divided by the sphere's cross-sectional area $\pi a^2$,
\begin{equation}
Q_\mathrm{abs,sca,ext} \equiv \frac{\sigma_\mathrm{abs,sca,ext}}{\pi a^2}.
\label{efficiencies}
\end{equation}

To obtain the cross-sections, we have to calculate their corresponding time-averaged powers in terms of the incident and scattered fields.
This involves the scalar product of two time-harmonic fields.
The time-average
over the wave period $2\pi/\omega$
 of two
time-harmonic functions $f_1\ee^{-\ii \omega t}$ and $f_2\ee^{-\ii \omega t}$, with
complex amplitudes $f_1$ and $f_2$, is given by
\begin{equation}
\label{f1f2}
\overline{f_1\ee^{-\ii \omega t} f_2\ee^{-\ii \omega t}}=\frac{1}{2}\re [f_1^* f_2],
\end{equation}
where  `$\re$' means the real-part of.

The total absorbed power is tantamount to minus the time-average of the radial total stress projected onto the element velocity $\partial_t \vec{u}$ and integrated over a control sphere of radius approaching to infinite,
\begin{equation}
\label{abs}
P_\textrm{abs} = - \lim_{r\rightarrow \infty}r^2 \oint_{4 \pi} \overline{\left(\partial_t \vec{u}\cdot{\bm \sigma}\right)}
\cdot\vec{e}_r\: \textrm{d} \Omega.
\end{equation}
For a time-harmonic displacement, the  element velocity is $\partial_t \vec{u}=-\ii \omega \vec{u}$.
Using the components of the displacement vector in spherical coordinates~\cite{Graff1991}
along with Eqs.~(\ref{uinL_far}), (\ref{uinMN_far}),
and (\ref{uscL_far}),
we find
\begin{subequations}
\begin{eqnarray}
u_r \nabla\cdot \vec{u}^* &=& u_r  \partial_r u_r^* +\bigO\left( r^{-3} \right), \\
\vec{u}\cdot (\nabla \vec{u}+
\nabla \vec{u}^{T})^* \cdot
\vec{e}_r&=&  2 u_r\partial_r u_r^*
+ u_\theta\partial_r u_\theta^*
+ u_\varphi\partial_r u_\varphi^* +
\bigO\left(r^{-3}\right).
\end{eqnarray}
\end{subequations}
Therefore, with these expressions and Eq.~(\ref{f1f2}),
we obtain the absorbed power  as
\begin{equation}
P_\mathrm{abs} = - \frac{\rho_0 \omega}{2} \lim_{r\rightarrow \infty}r^2 \re \oint_{4 \pi}
 \ii \bigl[  c_\LL^2  
 u_r \partial_r u_r^* 
+c_\SSS^2\left(
u_\theta\partial_r u_\theta^*
+ u_\varphi\partial_r u_\varphi^*\right)
 \bigr] \textrm{d} \Omega.
 \label{abs1}
\end{equation}
Referring to the asymptotic representation of the incident and scattered fields given in Eqs.~(\ref{uinL_far}), (\ref{uinMN_far}), and (\ref{uscL_far}), we may re-write Eq.~(\ref{abs1}) as
\begin{eqnarray}
\nonumber
&& P_\mathrm{abs} = 
 -  \frac{\rho_0 \omega u_0^2 }{2}  \lim_{r\rightarrow \infty}r^2
     \re \oint_{4 \pi}\ii \biggl[ 
c_\LL^2 
    \left(u_{r,\mathrm{in}}^{(\LL)}\partial_r u_{r,\mathrm{sc}}^{(\LL)\,*} 
  + u_{r,\mathrm{sc}}^{(\LL)}\partial_r u_{r,\mathrm{in}}^{(\LL)\,*}
  + u_{r,\mathrm{sc}}^{(\LL)}\partial_r u _{r,\mathrm{sc}}^{(\LL)\,*}\right)\\
 &+& 
  c_\SSS^2
    \left( \vec{u}_{\mathrm{sc}}^{(\SSS)}\cdot \partial_r \vec{u}_{\mathrm{in}}^{(\SSS)\,*}
   + \vec{u}_{\mathrm{in}}^{(\SSS)}\cdot \partial_r \vec{u}_{\mathrm{sc}}^{(\SSS)\,*}
  +
  \vec{u}_{\mathrm{sc}}^{(\SSS)}\cdot \partial_r \vec{u}_{\mathrm{sc}}^{(\SSS)\,*}
      \right)\,\biggr]\dd \Omega. 
    \label{abs3}
\end{eqnarray}
Importantly, 
the terms involving only the incident displacement vector $\vec{u}_\inc$
do not contribute to the absorbed power. 
Since they concern to the wave propagation without an inclusion,
we left them out in Eq.~(\ref{abs3}).
We recognize in Eq.~(\ref{abs3}) that terms involving only scattered fields are related to the scattering power
\begin{equation}
P_\mathrm{sca} = 
    \frac{\rho_0 u_0^2}{2} 
      \oint_{4 \pi} 
      \left[  c_\LL^3
      \left|\vec{f}^{(\LL)}(\theta, \varphi)
      \right|^2 +
       c_\SSS^3
            \left|\vec{f}^{(\SSS)}(\theta, \varphi)
            \right|^2\right] \dd\Omega.
    \label{sca}
\end{equation}

Now, we obtain
the absorption, scattering, and extinction cross-sections
in terms of the beam-shape and scattering coefficients.
Incorporating the expressions given in (\ref{power_components}) 
into Eq.~(\ref{abs3}) results
\begin{subequations}
\begin{eqnarray}
\label{sigma_absL}
\sigma_\mathrm{abs}^{(\LL)} 
&=&
-\frac{\rho_0 u_0^2 c_\LL^3}{2I_0}
  \re
   \sum_{n,m}
   \left(\left|s_{nm}^{(\LL)}\right|^2+
   s_{nm}^{(\LL)}a_{nm}^{(\LL)\,*}  \right), \\ 
\sigma_\mathrm{abs}^{(\SSS)} &=&   - \frac{\rho_0 u_0^2 c_\SSS^3}{2I_0}
         \re
           \sum_{n,m}
        n(n+1)
        \biggl(
        \left|s_{nm}^{(\SSS,1)}\right|^2 +\left|s_{nm}^{(\SSS,2)}\right|^2 
        +
        s_{nm}^{(\SSS,1)}a_{nm}^{(\SSS,1)\,*}
        +
        s_{nm}^{(\SSS,2)}a_{nm}^{(\SSS,2)\,*}
             \biggr).
             \label{sigma_absS}
\end{eqnarray}
\end{subequations}
The absorption cross-section is the sum of the longitudinal and shear components,
\begin{equation}\label{Sabs}
\sigma_\mathrm{abs} = \sigma_\mathrm{abs}^{(\LL)} + \sigma_\mathrm{abs}^{(\SSS)}.
\end{equation}
Similarly,
substituting the equations in (\ref{power_components})
into Eq.~(\ref{sca}), we get the longitudinal and shear scattering cross-section components,
\begin{subequations}
\begin{eqnarray}
\label{sigma_scaL}
\sigma_\mathrm{sca}^{(\LL)} 
&=& 
\frac{\rho_0 u_0^2 c_\LL^3}{2I_0}
   \sum_{n,m}
   \left|s_{nm}^{(\LL)}\right|^2, \\
\sigma_\mathrm{sca}^{(\SSS)}   
   &=&
  \frac{\rho_0 u_0^2 c_\SSS^3}{2I_0}
   \sum_{n,m}
   n(n+1)
           \left(
           \left|s_{nm}^{(\SSS,1)}\right|^2
           +\left|s_{nm}^{(\SSS,2)}\right|^2
                \right).
\label{sigma_scaS}
\end{eqnarray}
\end{subequations}
The scattering cross-section is then
\begin{equation}\label{Ssca}
\sigma_\mathrm{sca} = \sigma_\mathrm{sca}^{(\LL)} + \sigma_\mathrm{sca}^{(\SSS)}.
\end{equation}

Equations~(\ref{Sabs}) and (\ref{Ssca}) show that
the contribution of longitudinal and shear waves to
the absorption and scattering cross-sections are decoupled. 
Furthermore,
the extinction cross-section comes from the combination of these equations as follows
\begin{equation}
\sigma_\mathrm{ext} 
=-\frac{\rho_0 u_0^2}{2I_0}
  \re
   \sum_{n,m}
\biggl[
   c_\LL^3 s_{nm}^{(\LL)}   
      a_{nm}^{(\LL)\,*}
      + n(n+1) c_\SSS^3
           \left(
        s_{nm}^{(\SSS,1)}
         a_{nm}^{(\SSS,1)\,*}
        +
        s_{nm}^{(\SSS,2)}a_{nm}^{(\SSS,2)\,*}
             \right)
             \biggr].
             \label{sigma_ext2}
\end{equation}
This is \emph{the extended optical theorem} for elastic waves involving a spherical inclusion of a rigid, void, elastic, viscoelastic, or layered material.
The properties of the inclusion appear in the scattering coefficients $s_{nm}^{(\LL)}$,  $s_{nm}^{(\SSS,1)}$ and $s_{nm}^{(\SSS,2)}$, while the beam characteristics are present in the beam-shape coefficients
$a_{nm}^{(\LL)}$, $a_{nm}^{(\SSS,1)}$, and $a_{nm}^{(\SSS,2)}$.
The extinction power
 of the longitudinal or shear wave
 can only happen if that component
 is present in the incident wave.
 For instance, if the incident wave  is a shear wave only then the longitudinal extinction
 cross-section is zero, $\sigma_\mathrm{ext}^{(\LL)}=0$.
 Even though longitudinal scattered waves are present in the medium due to mode conversion.
 Same thing happens when the incident wave is purely longitudinal, 
 $\sigma_\mathrm{ext}^{(\SSS)}=0$.

\subsection{Boundary conditions and coefficient relations}
In the scattering by an isotropic solid sphere embedded in an elastic solid matrix, the
boundary conditions require the continuity of the displacement vectors and the stress tensor across the sphere's surface $r = a$.
On assuming that the inclusion does not have an energy source, the absorption cross-section
satisfies $\sigma_\mathrm{abs} \ge 0$.
This fact will be used to establish the relations that should be satisfied by the scattering coefficients.
Let us consider a longitudinal and shear incident wave separately.

\subsubsection{Longitudinal waves} 

Longitudinal waves are characterized by the beam-shape coefficient $a_{nm}^{(\LL)}$.
We see from the equations~\eqref{Hensen_vectors},
only shear scattered waves of the second-type described by $s_{nm}^{(\SSS,2)}$ can be produced by mode conversion.
Defining $s_{nm}^{(\LL)}= s_n^{(\LL)} a_{nm}^{(\LL)}$ and $s_{nm}^{(\mathrm{S},2)}= s_{n}^{(\SSS,2)} a_{nm}^{(\LL)}$; while
the transmission coefficients are 
$t_{nm}^{(\LL)}= t_n^{(\LL)} a_{nm}^{(\LL)}$ and
$t_{nm}^{(\mathrm{S},2)}= t_{n}^{(\SSS,2)} a_{nm}^{(\LL)}$.
The four unknown coefficients $s_n^{(\LL)},s_n^{(\SSS,2)},t_n^{(\LL)},$ and $t_n^{(\SSS,2)}$ are determined from the following boundary conditions
\begin{equation}
\oint_{4 \pi}\left[
\begin{matrix}
\left(
\vec{u}_\inc + \vec{u}_\sca - \vec{u}_\tra\right) \cdot
\vec{Y}_{nm}^*\\
\left(\vec{u}_\inc + \vec{u}_\sca - \vec{u}_\tra\right) \cdot
 \vec{\Psi}_{nm}^*\\
 \vec{e}_r
 \cdot \left(\vec{\sigma}_{\inc}  + \vec{\sigma}_{\sca}  -
 \vec{\sigma}_{\tra} \right) \cdot
 \vec{Y}_{nm}^*\\
 \vec{e}_r
  \cdot \left(\vec{\sigma}_{\inc}  + \vec{\sigma}_{\sca}  -
  \vec{\sigma}_{\tra} \right) \cdot
  \vec{\Psi}_{nm}^*
\end{matrix}
\right]\dd \Omega = 0.
\label{boundary_L}
\end{equation}
From Eqs.~(\ref{sigma_absL}) and (\ref{sigma_absS}), and knowing that $\sigma_\mathrm{abs}\ge 0$, we find that the scaled scattering coefficients satisfy
\begin{equation}
\label{longitudinal_ineq}
   \re\left [s_n^{(\LL)}\right] +
\left|s_{n}^{(\LL)}\right|^2+
   n(n+1)\left(\frac{c_\SSS}{c_\LL}\right)^3\left|s_{n}^{(\SSS,2)}\right|^2 \le 0, \quad n=0,1,2,\dots
\end{equation}

\subsubsection{Shear waves}
Shear waves are described by two beam-shape coefficients, namely, $a_{nm}^{(\SSS,1)}$ and $a_{nm}^{(\SSS,2)}$.
Longitudinal waves are produced from shear waves of the second-type
through mode conversion.
Thus,  we rewrite the scattering and transmission coefficients as $s_{nm}^{(\mathrm{S},j)}= s_{n}^{(\SSS,j)} a_{nm}^{(\mathrm{S},j)}$ 
$(j=1,2)$, $s_{nm}^{(\LL)}= s_n^{(\LL)} a_{nm}^{(\mathrm{S},2)}$; and
$t_{nm}^{(\mathrm{S},j)}= t_{n}^{(\SSS,j)} a_{nm}^{(\mathrm{S},j)}$ 
$(j=1,2)$, $t_{nm}^{(\LL)}= t_n^{(\LL)} a_{nm}^{(\mathrm{S},2)}$.
We need six conditions to determine the unknown coefficients
$s_n^{(\LL)}$, $t_n^{(\LL)}$,
$s_n^{(\SSS,j)}$ and $t_n^{(\SSS,j)}$, with $j=1,2$.
Four conditions are already given in Eq.~(\ref{boundary_L}), whereas the additional conditions are
\begin{equation}
\oint_{4 \pi}\left[
\begin{matrix}
\left(
\vec{u}_\inc + \vec{u}_\sca - \vec{u}_\tra\right) \cdot
\vec{\Phi}_{nm}^*\\
 \vec{e}_r
 \cdot \left(\vec{\sigma}_{\inc}  + \vec{\sigma}_{\sca}  -
 \vec{\sigma}_{\tra} \right) \cdot
 \vec{\Phi}_{nm}^*
\end{matrix}
\right]\dd \Omega = 0.
\label{boundary_S}
\end{equation}
Having that $\sigma_\mathrm{abs} \ge 0$ and using Eqs.~(\ref{sigma_absL}) and (\ref{sigma_absS}), we obtain the following relations for the scaled scattering coefficients
\begin{equation}
\label{shear_ineq}
\re\left[s_{n}^{(\SSS,1)}+
s_{n}^{(\SSS,2)}\right] +
\left|s_{n}^{(\SSS,1)}\right|^2 
+
        \left|s_{n}^{(\SSS,2)}\right|^2 
              + \frac{1}{n(n+1)}\left(\frac{c_\LL}{c_\SSS}\right)^3\left|s_{n}^{(\LL)}\right|^2\le 0, \quad n=1,2,3,\dots
\end{equation}
The relations in (\ref{longitudinal_ineq}) and (\ref{shear_ineq}) are a necessary condition to be satisfied by the scaled scattering coefficients.

\subsection{Elastic radiation force}
In this section, we use the optical theorem formalism to derive the elastic radiation force 
exerted by traveling plane waves propagating along the $z$-axis on a spherical inclusion of radius $a$.
The linear momentum density $p$ carried by a plane wave is related to the
time-averaged energy density $E_0$   through the expression~\cite[p.~234]{Elmore1985} 
\begin{equation}
E_0 = 
\begin{cases}
p c_{\LL}, \quad \mathrm{longitudinal}\\
p c_{\SSS},\quad \mathrm{shear}.
\end{cases}
\label{energy-momentum}
\end{equation}
Meanwhile the time-averaged force per unit area (radiation pressure) exerted on
a nonreflective interface due to  linear momentum transfer from the incident wave is also~\cite{Torr1984}
$ p c_{\LL,\SSS} $.
From the energy-momentum
relation in Eq.~(\ref{energy-momentum}), we conclude that the change in the linear momentum of the incident beam in the scattering process is proportional to extinction power $P_\mathrm{ext}=P_\mathrm{abs}+P_\mathrm{sca}$ and thus related to the cross-section $\sigma_\mathrm{ext}=\sigma_\mathrm{abs}+\sigma_\mathrm{sca}$.
The linear momentum change corresponding to absorption $\sigma_\mathrm{abs}$ cannot be not replaced.
In contrast, the part relative to the scattered power returns to the medium.
The projected linear momentum of the scattered waves on the forward direction $\theta=0^\circ$ should be calculated
to obtain the elastic radiation force.

As the scattered wave approaches the farfield region $k_\LL r, k_\SSS r\gg 1$, it resembles 
a traveling plane wave. 
Thus, according to Eq.~(\ref{energy-momentum}),  
the scattered  linear momentum density 
in  an arbitrary direction at the farfield  is given by
\begin{equation}
\bm{p}_r = \frac{E_0}{c_j r^2}\frac{\dd \sigma_\mathrm{sca}}{\dd \Omega} \,\vec{e}_r, \quad j\in\{\LL,\SSS\},
\end{equation}
where
\begin{equation}
\frac{\dd \sigma_\mathrm{sca}}{\dd \Omega}
=  
  \frac{\rho_0 u_0^2}{2 I_0} 
        \left[  c_\LL^3
      \left|\vec{f}^{(\LL)}(\theta, \varphi)
      \right|^2 +
       c_\SSS^3
            \left|\vec{f}^{(\SSS)}(\theta, \varphi)
            \right|^2\right]
\end{equation} 
is the differential scattering cross-section, which follows from Eq.~(\ref{sca}).
Note that the scattering cross-section is
$
\sigma_\mathrm{sca}=r^{-2}\oint_{4 \pi} (\dd \sigma_\mathrm{sca}/\dd \Omega) r^2 \dd\Omega.$
The total linear momentum density along the forward direction is just
\begin{equation}
\label{forward_momentum}
\oint_{4\pi} (\bm{p}_r\cdot \bm{e}_z)\, r^2 \dd\Omega = \frac{E_0}{c_j}
\oint_{4\pi} \cos \theta \frac{\dd \sigma_\mathrm{sca}}{\dd \Omega} \dd\Omega.
\end{equation}
We recognize that the right-hand side of this equation is related to the spatial average of cosine of the polar scattering angle, also known as the asymmetry parameter~\cite[p.~72]{Boheren1998},
\begin{equation}
\langle\cos \theta\rangle
=  \frac{1}{
\sigma_\mathrm{sca}}
\oint_{4 \pi}
\cos \theta 
\frac{\dd \sigma_\mathrm{sca}}{\dd \Omega}\dd \Omega.
\label{costheta}
\end{equation}
For symmetric scattering about $\theta=90^\circ$, the asymmetry parameter is zero, $\langle \cos \theta \rangle = 0$.
If the scattering is more prominent in the forward direction ($\theta=0^\circ$), $\langle \cos \theta \rangle$ is positive.
The asymmetry parameter is negative when more scattering occurs toward
the backward direction $\theta = 180^\circ$.
By using the relation $E_0=I_0/c_j$ in Eq.~{\eqref{forward_momentum}}, the total linear momentum along the axial direction  is
given by
$(I_0/c_j^2)
\langle\cos \theta\rangle\,
\sigma_\mathrm{sca}.$

Now we can state that
the elastic radiation force 
is given in terms of the linear momentum extracted from the incident wave and that part taken away by the scattered waves in the forward direction $(\theta = 0^\circ)$,
\begin{equation}
F^{(j)}_\rad =   \frac{I_0}{c_j} \sigma_\rad,\quad j\in\{\LL,\SSS\}
\label{RF2}
\end{equation}  
where 
 \begin{equation}
 \label{sigma_p}
 \sigma_\mathrm{rad} =
 \sigma_\mathrm{ext} - \,
 \langle\cos \theta\rangle\,
 \sigma_\mathrm{sca} 
 =
 \sigma_\mathrm{abs} + 
 \left(1- \,
  \langle\cos \theta\rangle\right)
  \sigma_\mathrm{sca} 
 \end{equation} 
is the radiation force cross-secion.
Thus, the radiation force efficiency reads
 \begin{equation}
  \label{Q_rad}
  Q_\mathrm{rad} 
  =
  Q_\mathrm{abs} + 
  \left(1- \,
   \langle\cos \theta\rangle\right)
   Q_\mathrm{sca}.
  \end{equation}
In terms of this efficiency we have
\begin{equation}
F^{(j)}_\rad =  \pi a^2 \frac{I_0}{c_j} Q_\rad.
\label{RF}
\end{equation}  
For a non-absorbing inclusion, the elastic radiation force depends only on the scattering efficiency, 
\begin{equation}
 F^{(j)}_\rad = \pi a^2
 \frac{I_0}{c_j}
  \left(1- \,
   \langle\cos \theta\rangle \right)
   Q_\mathrm{sca}.
 \end{equation}
 
To obtain a useful radiation force formula, we need to calculate 
$Q_\mathrm{sca} \langle \cos \theta \rangle$ in terms of the scattering coefficients.
In so doing, we use Eqs.~(\ref{fL}) and (\ref{fS}) with (\ref{app:ortho}) and (\ref{app:recurrence}) into  Eq.~(\ref{costheta}).
 Accordingly, we find
 \begin{eqnarray}
 \nonumber
&&\langle\cos \theta\rangle Q_\mathrm{sca}
 = \\
 \nonumber
 && 
 \frac{\rho_0  u_0^2 }{\pi a^2 I_0}
 \re
 \sum_{n,m} \biggl\{ \ii 
 \sqrt{ \frac{(n+m+1)(n-m+1)}{(2n + 3) (2n + 1)} }
 \left[c_\LL^3
 s_{n+1,m}^{(\LL)*}s_{nm}^{(\LL)}
 + c_\SSS^3  n(n+2) 
 \left(
s_{n+1,m}^{(\SSS,2)*}s_{nm}^{(\SSS,2)}
+  s_{n+1,m}^{(\SSS,1)*}s_{nm}^{(\SSS,1)}
 \right)\right]\\
 &&
 + c_\SSS^3 m s_{nm}^{(\SSS,1)}s_{nm}^{(\SSS,2)*}
 \biggr\}.
 \label{avg_cos_theta}
 \end{eqnarray}
Here, the contribution of the longitudinal and shear scattered waves are decoupled.
However, the last term within the curly brackets involves a crossed contribution of both types of shear scattered waves.
 
Let us examine the direction  of the elastic radiation force.
Granted that no energy source is inside the inclusion,
the absorption efficiency satisfies $Q_\mathrm{abs}\ge 0$.
Moreover, the scattering efficiency is always positive, $Q_\mathrm{sca}> 0$.
So, having $|\langle \cos \theta \rangle | < 1$, we conclude that
\begin{equation}
\label{ineq_cos_theta}
0< \left(1- \langle\cos \theta\rangle \right)Q_\mathrm{sca}< 2 Q_\mathrm{sca}.
\end{equation}
Thus, the radiation force efficiency satisfies 
\begin{equation}
0<Q_\mathrm{rad}< Q_\mathrm{sca} + Q_\mathrm{ext}.
\end{equation}
This implies that
the elastic radiation force due to a traveling plane wave always points towards to the forward scattering direction.
For a non-dissipative inclusion, we have $0<Q_\mathrm{rad}<2 Q_\mathrm{sca}$.

A similar result to Eq.~(\ref{RF}) has been earlier obtained for the acoustic radiation force in  fluids caused by  a plane 
wave~\cite{Westervelt1951,Olsen1958} and a Bessel beam~\cite{Zhang2011}.
The same expression has also been found for electromagnetic plane waves~\cite{Hulst1981}.
This shows an universal character of the radiation force phenomenon.


\section{Some wave examples}

\subsection{Longitudinal plane wave }
Consider an incident longitudinal plane  wave (LPW) propagating along the $z$-axis toward infinity.
The corresponding displacement vector is $\vec{u}_\mathrm{in}=u_0
\vec{u}_\mathrm{in}^{(\LL)} \ee^{-\ii \omega t}$, with its amplitude being
\begin{equation}
\vec{u}_\mathrm{in}^{(\LL)}  =
 -\ii u_0
\nabla_\LL \ee^{\ii k_\LL z}.
 \label{uin_plane}
\end{equation}
The time-averaged incident intensity is 
\begin{equation}
\label{I0L}
I_0=-
 \overline{\partial_t \bm{u}_\inc \cdot \bm{\sigma} \cdot \bm{e}_z }.  
\end{equation}
From Eqs.~(\ref{stress-strain}) and (\ref{uin_plane}), we find $\vec{\sigma}\cdot \vec{e}_z=\ii \omega \rho_0 c_\LL \vec{u}_{\inc}$
and $\partial_t \vec{u}_\inc = -\ii \omega \vec{u}_\inc$.
Inserting these expressions into Eq.~(\ref{I0L}) yields
\begin{equation}
\label{I0LPW}
I_0=\frac{\rho_0 c_\LL (\omega u_0)^2}{2}.
\end{equation}
The partial-wave expansion of the incident displacement vector is given by~\cite{Brill1987}
\begin{eqnarray}
\nonumber
\vec{u}_\inc^{(\LL)} &=& -
\sum_{n=0}^\infty 
\ii^{n+1}  (2n+1)\,  \nabla_\LL \left[
j_n(k_\LL r) P_{n}(\cos \theta)\right]\\
&=& -
\sum_{n=0}^\infty 
\ii^{n+1}  \sqrt{4 \pi (2 n+1)}  \vec{L}^{(1)}_{n,0}.
\end{eqnarray}
Referring to Eqs.~(\ref{uinL}) and (\ref{SH}), the longitudinal beam-shape coefficient reads
\begin{equation}
\label{anmLpw}
a_{nm}^{(\LL)}= - \ii^{n+1} \sqrt{4 \pi (2 n+1)}  \delta_{m,0}.
\end{equation}
Because the vector spherical harmonics in the equations of (\ref{Hensen_vectors}) are orthogonal,
the longitudinal-to-shear mode conversion in the scattering process  only involves
the second scattering coefficient $s_{nm}^{(\SSS,2)}$.
Thus, the scattered displacement vector is given by
\begin{eqnarray}
\nonumber
\vec{u}_\sca &=& -
\sum_{n=0}^\infty 
\ii^{n+1}  (2n+1)\,  \left[ s_n^{(\LL)}\nabla_\LL [h_n^{(1)}(k_\LL r)P_{n}(\cos \theta)] + s_n^{(\SSS,2)}
\nabla_\SSS \times \nabla_\SSS \times [k_\SSS r h_n^{(1)}(k_\SSS r) P_{n}(\cos \theta) \vec{e}_r] \right]
 \\
 &=&
 \sum_{n=0}^\infty s_{n,0}^{(\LL)} \vec{L}^{(2)}_{n,0} +
 s_{n,0}^{(\SSS,2)} \vec{N}^{(2)}_{n,0}.
\end{eqnarray}
We readily recognize that
the longitudinal and shear scattering coefficients are expressed by
\begin{equation}
\label{snmL_longitudinal}
\left(
\begin{matrix}
s_{nm}^{(\LL)}\\
s_{nm}^{(\SSS,2)}
\end{matrix}
\right)
 = -\ii^{n+1} \sqrt{4 \pi (2 n+1)} \delta_{m,0} 
\left(
\begin{matrix} 
 s_{n}^{(\LL)} \\
 s_{n}^{(\SSS,2)}
 \end{matrix}
 \right),
\end{equation}
where $s_n^{(\LL)}$ and $s_n^{(\SSS,2)}$ are coefficients given in~\eqref{scaled_scatt_coeff}.
We also note that they related by 
\begin{equation}
\label{sS2sL}
s_n^{(\SSS,2)}= \frac{c_\LL}{c_\SSS} \frac{\det \mathbf{D}_n^{(\SSS)}}{\det \mathbf{D}_n^{(\LL)}} s_n^{(\LL)},
\end{equation}
where $\mathbf{D}_n^{(\LL)}$ and $\mathbf{D}_n^{(\SSS)}$ are matrices given in \ref{app:scatt_coefficients}.

By substituting Eqs.~(\ref{anmLpw}) and (\ref{snmL_longitudinal}) into Eqs.~(\ref{sigma_absL}) and (\ref{sigma_absS}),
we obtain the absorbing and scattering efficiencies  as
\begin{subequations}
\begin{eqnarray}
Q_\mathrm{abs}^\LPW &=& -
\frac{4 }{x_\LL ^2}
  \re
   \sum_{n=0}^\infty
   \left(2n+1\right) \left[
   s_{n}^{(\LL)} + \left|s_{n}^{(\LL)}\right|^2
   + n(n+1)\left(\frac{c_\SSS}{c_\LL}\right)^3  \left|s_{n}^{(\SSS,2)}\right|^2
   \right],
   \label{sigma_abs_longitudinal}
   \\
\label{sigma_sc_pw}
Q_\mathrm{sca}^\LPW &=& 
\frac{4}{x_\LL ^2}
   \sum_{n=0}^\infty
   \left(2n+1\right)
   \left[
       \left|s_{n}^{(\LL)}\right|^2
       +
       n(n+1)\left(\frac{c_\SSS}{c_\LL}\right)^3
           \left|s_{n}^{(\SSS,2)}\right|^2
           \right],
\end{eqnarray}
\end{subequations}
where $x_\LL=k_\LL a$ is the longitudinal sphere size parameter.
Note that the shear scattering coefficient $s_n^{(\SSS,2)}$ can be eliminated in the absorption
and scattering efficiencies by means of  Eq.~\eqref{sS2sL}.
The scattering cross-section presented here 
 agrees with that previously obtained in Ref.~\cite{Ying1956}.
One can verify this
by setting 
$
s_n^{(\LL)} = \ii^{-n-1} k_\LL A_n^*/(2n+1)$ and
$s_n^{(\SSS,2)} = -\ii^{-n-1} k_\SSS B_n^*/(2n+1)$,
where $A_n$ and $B_n$ are the expansion coefficients in the notation of Ref.~\cite{Ying1956}.

Referring to Eq.~(\ref{fL_forward}), the forward scattering function is given 
by
\begin{equation}
\vec{f}^{(\LL)}(0,0) =-\ii
\sum_{n=0}^\infty
   \left(2n+1\right)
   s_{n}^{(\LL)} \vec{e}_z.
\end{equation}
Hence, using Eq.~(\ref{sigma_ext2}), we find
the optical theorem 
\begin{equation}
\sigma_\mathrm{ext}^\LPW =
   \frac{4 \pi}{k_\LL^2} \im \left[\vec{f}^{(\LL)}(0,0)\right]\cdot \vec{e}_z,
   \label{sigma_ext_pwL}  
\end{equation}
where `Im' means the imaginary-part of.
This equation states that the extinction cross-section is related to the scattering  function
along the the forward  direction~\cite{Barratt1965}.

Now we are able to calculate the efficiency  of the elastic radiation force exerted
on the inclusion.
The linear momentum taken away from the incident wave
is obtained substituting the scattering coefficients given in Eq.~(\ref{snmL_longitudinal}) into 
Eq.~(\ref{avg_cos_theta}).
Finally, using Eqs.~(\ref{sigma_abs_longitudinal}) and (\ref{sigma_sc_pw}), we arrive at
\begin{equation}
Q_\mathrm{rad}^\LPW =-
\frac{4 }{x_\LL^2  }      \re
      \sum_{n=0}^\infty \biggl[
      \left(2n+1\right)
      s_{n}^{(\LL)}
      + 2 (n+1) 
      \left\{
      s_{n}^{(\LL)} s_{n+1}^{(\LL)*}
      +n (n+2)\left(\frac{c_\SSS}{c_\LL}\right)^3
      s_{n}^{(\SSS,2)} s_{n+1}^{(\SSS,2)*}
      \right\} \biggr].
   \label{RF_plane_longitudinal}
\end{equation}
As previously noted, the coefficient $s_n^{(\SSS,2)}$ can be eliminated through
the relation in Eq.~\eqref{sS2sL}.
The last term in the curly brackets are due to mode conversion. 

\subsection{Shear plane wave}

Without loss of generality, we assume that 
the shear plane  wave (SPW) is polarized along the $x$-axis and propagates on the $z$-axis
toward infinity.
Thus, the incident displacement vector reads
\begin{equation}
\vec{u}_\mathrm{in} = u_0
\vec{u}_\mathrm{in}^{(\SSS)} = -\ii
u_0\ee^{\ii k_\SSS z}\vec{e}_x.
\label{uin_shear}
\end{equation}
Referring to Eq.~(\ref{I0L}) 
and noting that $\vec{\sigma}\cdot \vec{e}_z = \ii \omega \rho_0 c_\SSS \vec{u}_\inc $,
we attain
the  intensity magnitude
\begin{equation}
I_0=\frac{\rho_0  c_\SSS (\omega u_0)^2}{2}.
\label{I0SPW}
\end{equation}
The partial-wave expansion
of the shear plane wave is given by~\cite{Brill1987}
\begin{equation}
\vec{u}_\inc^{(\SSS)}
=
 \sum_{n=1}^\infty
\ii^{n+1}\frac{ 2n+1}{n(n+1)}
\nabla_\SSS \times 
\left[\left( \ii\sin \varphi + \nabla_\SSS
\times \cos \varphi  
\right)
   k_\SSS r\, j_n(k_\mathrm{s} r)
  P_{n}^1(\cos \theta)  \vec{e}_r\right].  
  \label{uSPW}
\end{equation}
To obtain the beam-shape coefficients $a_{nm}^{(\SSS,1)}$ and $a_{nm}^{(\SSS,2)}$ we first note that
\begin{equation}
Y_n^{\pm 1}(\theta,\varphi) = \pm
\sqrt{\frac{ 2n+1}{4\pi n(n+1)} }P_{n}^{1}(\cos \theta) \,\ee^{\pm \ii  \varphi}.
\end{equation}
Thus,
\begin{equation}
\left(
\begin{matrix}
\sin \varphi\\
-\ii \cos \varphi
\end{matrix} 
\right)
P_{n}^{1}(\cos \theta) \vec{e}_r = \frac{1}{2 \ii}\sqrt{\frac{4\pi n(n+1) }{2n+1} } 
\left(
\begin{matrix}
\vec{Y}_{n,1}(\theta,\varphi) + \vec{Y}_{n,-1}(\theta,\varphi)\\
\vec{Y}_{n,1}(\theta,\varphi) - \vec{Y}_{n,-1}(\theta,\varphi)
\end{matrix}
\right)
\end{equation}
From Eqs.~(\ref{uinMN}), (\ref{M1}), and (\ref{N1}), we rewrite the expansion in Eq.~(\ref{uSPW}) as
\begin{equation}
\vec{u}_\inc^{(\SSS)}
=
\sum_{n=1}^\infty
\frac{\ii^{n+1}}{2}\sqrt{4 \pi \frac{2n+1}{n (n+ 1)}}
\left(
\vec{M}_{n,1}^{(1)} + \vec{M}_{n,-1}^{(1)} + 
\vec{N}_{n,1}^{(1)} - \vec{N}_{n,-1}^{(1)}
\right).
  \label{uSPW2}
\end{equation}
Hence,
we find that the beam-shape coefficients be
expressed by
\begin{equation}
\left(
\begin{matrix}
 a_{nm}^{(\SSS,1)}\\
 a_{nm}^{(\SSS,2)}
\end{matrix}
 \right) 
= 
\frac{  \ii^{n+1} }{2}  \delta_{m,\pm 1}
\sqrt{4 \pi \frac{2n+1}{n (n+ 1)}}
\left(
\begin{matrix}
1\\
m
\end{matrix}
 \right).
\label{shear_anm}
\end{equation}

In the scattering process, shear-to-longitudinal mode conversion takes place. 
However,  shear waves described by the vector spherical harmonic $\vec{\Phi}_{nm}$ cannot be
converted to a longitudinal wave.
Hence, the beam-shape coefficient $a_{nm}^{(\SSS,1)}$ cannot be associated to the longitudinal scattered wave.
Consequently, this wave should have $\cos \varphi$ dependence.
Thus, we may express the scattered waves as
\begin{eqnarray}
\nonumber
\vec{u}_\sca
&=& 
\sum_{n=1}^\infty
\ii^{n+1}\frac{ 2n+1}{n(n+1)}\biggl[ 
\nabla_\SSS \times \left(\ii s_{n}^{(\SSS,1)} \sin \varphi + s_{n}^{(\SSS,2)}\nabla_\SSS
\times \cos \varphi 
\right)
   k_\SSS r h_n^{(1)}(k_\mathrm{S} r)   P_{n}^1(\cos \theta) \vec{e}_r\\
   &+&
   s_{n}^{(\LL)}
   \nabla_\LL \left(\cos \varphi\, h_n^{(1)}(k_\mathrm{L} r)   P_{n}^1(\cos \theta) \right) 
    \biggr]
    \nonumber
    \\
&=& \sum_{n=1}^\infty \sum_{m=-1}^1
\left(
s_{nm}^{(\SSS,1)}
\vec{M}_{n,m}^{(2)}  
+ 
 s_{nm}^{(\SSS,2)} \vec{N}_{n,m}^{(2)} 
+
 s_{nm}^{(\LL)} \vec{L}_{n,m}^{(2)}
\right).
\end{eqnarray}
The scattering coefficients are given by
\begin{equation}
\label{SW_scatt_coeff}
\left(
\begin{matrix}
s_{nm}^{(\SSS,1)}\\
s_{nm}^{(\SSS,2)}\\
s_{nm}^{(\LL)}
\end{matrix}
\right)
 = 
\frac{  \ii^{n+1} }{2} \delta_{m,\pm 1}
\sqrt{4 \pi \frac{(2n+1)}{n (n+ 1)}} 
\left(
\begin{matrix}
  s_{n}^{(\SSS,1)}\\
ms_{n}^{(\SSS,2)}\\
ms_n^{(\LL)}
\end{matrix}
\right).
\end{equation}
where $s_n^{(\SSS,1)}$, $s_n^{(\SSS,2)}$, and $s_n^{(\LL)}$ are  obtained from 
the boundary conditions across the inclusions' surface. 
They are given in~\eqref{scaled_scatt_coeff2}.

Now we can obtain the efficiencies
by substituting 
the scattering coefficients into Eqs.~(\ref{sigma_absL}),  (\ref{sigma_absS}),
(\ref{sigma_scaL}), and (\ref{sigma_scaS}), we arrive at
\begin{subequations}
\begin{eqnarray}
Q_\mathrm{abs}^\SPW
   &=& -
        \frac{ 2 }{x_\SSS^2}
           \sum_{n=1}^\infty
        (2n+1)
        \biggl[\re
        \left(s_{n}^{(\SSS,1)}
                + s_{n}^{(\SSS,2)}\right)
                +
        \left|s_{n}^{(\SSS,1)}\right|^2 +  \left|s_{n}^{(\SSS,2)}\right|^2
        +\frac{1}{n(n+1)}\left(
        \frac{c_\LL}{c_\SSS}\right)^3
        \left|s_{n}^{(\LL)}\right|^2
        \biggr],
             \label{sigma_absS2}\\
Q_\mathrm{sca}^\SPW &=  &
                     \frac{ 2 }{x_\SSS^2}
                        \sum_{n=1}^\infty
                     (2n+1)
                     \biggl[
                     \left|s_{n}^{(\SSS,1)}\right|^2
                     +\left|s_{n}^{(\SSS,2)}\right|^2
                     +\frac{1}{n(n+1)}\left(
                             \frac{c_\LL}{c_\SSS}\right)^3 \left|s_{n}^{(\LL)}\right|^2
                             \biggr],
                          \label{sigma_scaS2}                          
\end{eqnarray}
\end{subequations}
where $x_\SSS=k_\SSS a$ is the shear size parameter of the sphere.
The scattering efficiency agrees with the result presented in~\cite{Einspruch1960}.
Inserting the coefficients given in Eq.~(\ref{SW_scatt_coeff}) into Eq.~(\ref{fS_forward}) yields the forward scattering function
\begin{equation}
\vec{f}^{(\SSS)}(0,0)= -
\frac{\ii}{2} \sum_{n=1}^\infty  
(2n+1)\,
 \left(s_{n}^{(\SSS,1)}+s_{n}^{(\SSS,2)} \right) \vec{e}_x .
\label{fSex}
 \end{equation}
Substituting the scattering coefficients given in (\ref{SW_scatt_coeff}) into Eq.~(\ref{sigma_ext2}), 
we obtain the optical theorem 
\begin{equation}
\sigma_\mathrm{ext}^\SPW = 
\frac{4 \pi}{k_\SSS^2}
\im\left[
\vec{f}^{(\SSS)}(0,0)\right]\cdot\vec{e}_x.
\end{equation}
The extinction cross-section depends on the projection of the scattering function onto the polarization 
direction.
This is in agreement with previous derivations~\cite{Barratt1965}.

Finally, we obtain the efficiency of the elastic radiation force 
by using the scattering coefficients from (\ref{SW_scatt_coeff}) into
Eqs.~(\ref{sigma_p}). 
Accordingly, we find 
\begin{eqnarray}
Q_\mathrm{rad}^\SPW =&
-& \frac{2 }{x_\SSS^2 }
\re\sum_{n=1}^\infty 
\biggl[ 
(2n+1)\left(
        s_{n}^{(\SSS,1)}
        +s_{n}^{(\SSS,2)}
        \right) 
 + \frac{2}{n+1} \biggl\{n(n+2)\biggl(s_{n}^{(\SSS,1)}s_{n+1}^{(\SSS,1)*}
  +  s_{n}^{(\SSS,2)}s_{n+1}^{(\SSS,2)*}\biggr)\nonumber\\
  &+&\frac{2n+1}{n}
  s_{n}^{(\SSS,1)}s_{n}^{(\SSS,2)*}
  +
 \left(\frac{{c}_{\LL}}{c_\SSS}\right)^3
 s_{n}^{(\LL)}  s_{n+1}^{(\LL)*}\biggr\}
 \biggr].
   \label{RF_plane_shear}
\end{eqnarray}
The last term in this equation is due to the shear-to-longitudinal mode conversion
in the scattering process.
The crossing term $s_{n}^{(\SSS,1)}s_{n}^{(\SSS,2)*}$ shows that the contribution from the scattered shear waves
is not decoupled.

\subsection{Longitudinal focused beam}
\begin{figure}
\centering
\includegraphics[scale=.55]{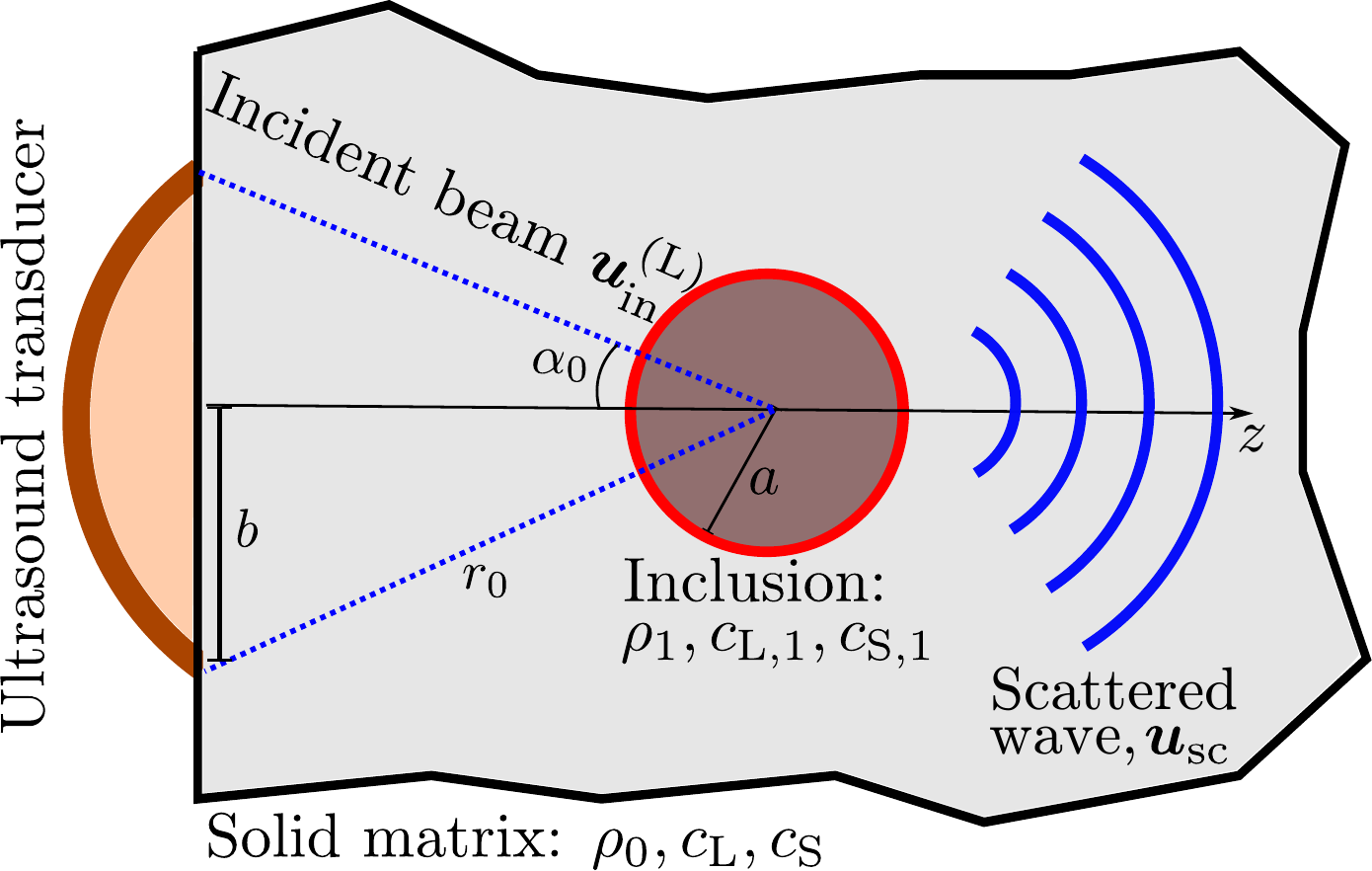} 
\caption{\label{fig:scatt_focused}
(Color online) The scattering of a longitudinal focused  beam by a sphere of radius $a$.
The incident beam is produced by a spherically focused transducer of aperture $2b$ and 
half-aperture angle $\alpha_0 = \arcsin(b/r_0)$.
}
\end{figure}

We assume that an spherically focused transducer produces longitudinal waves that are scattered by the spherical inclusion
placed at the transducer focus (on-focus configuration).
The transducer has aperture $2b$, radius of curvature $r_0$, and half-spread angle $\alpha_0=\arcsin(b/r_0)$ 
as depicted in Fig.~\ref{fig:scatt_focused}.
We consider the transducer in the paraxial approximation, where
its aperture is much  larger than the wavelength $k_\LL b\gg1$.
This implies that $k_\LL r_0\gg 1$.
In a lossless medium, the axial incident pressure yielded by the transducer is given by~\cite{Lucas1982}
\begin{equation}
p_\text{in} =  \frac{\ii p_0 r_0 \ee^{\ii k_\LL z}}{z - r_0} \left( 1 - \exp
 \left[\frac{\ii k_\LL b^2}{2}\left(\frac{1}{z}-\frac{1}{r_0}\right) \right]\right),
\end{equation}
where $p_0$ is the pressure magnitude at the source.
It is useful to  normalize the focused pressure by its maximum value at $z=r_0$, i.e. $|p_\mathrm{in}(r_0)|=k_\LL b^2 p_0/(2 r_0)$. 
One can show that the  beam-shape coefficient of the focused beam is given by~\cite{Silva2015}
\begin{equation}
a_{nm}^{(\LL)} = -\ii^n \sqrt{4\pi (2n + 1)} \delta_{m,0} \, g_n,
\end{equation}
where
\begin{equation}
 g_n =
\frac{2 \ee^{\ii k_\LL r_0}}{(2 n + 1) \sin^2 \alpha_0} \left[ P_{n+1}(\cos \alpha_0) - P_{n-1}(\cos \alpha_0) \right]
\end{equation}
is the diffraction coefficient.
The weakly focused limit $r_0\gg b$ leads to
\begin{equation}
 g_n = -  \ee^{\ii k_\LL r_0} +
 \bigO\left[\left(\frac{b}{r_0}\right)^2\right],
\end{equation}
which corresponds to the plane wave limit.

Referring to Eqs.~(\ref{snmL_longitudinal}), the longitudinal and shear scattering coefficients
are expressed by
\begin{equation}
\left(
\begin{matrix}
s_{nm}^{(\LL)} \\
s_{nm}^{(\SSS)} 
\end{matrix}
\right) =  - \ii^n \sqrt{4\pi (2n + 1)} \delta_{m,0} \, g_n
\left(
\begin{matrix}
s_{n}^{(\LL)} \\
s_{n}^{(\SSS,2)} 
\end{matrix}\right).
\label{sca_coeff_focused}
\end{equation}
The coefficients $s_{n}^{(\LL)}$ and $s_{n}^{(\SSS,2)}$ are the same as those found in the scattering of a longitudinal 
plane wave, since the inclusion is isotropic -- see the equations in~\eqref{scaled_scatt_coeff}.

After inserting Eq.~(\ref{sca_coeff_focused})  into Eqs.~(\ref{sigma_absL}),  (\ref{sigma_absS}),
(\ref{sigma_scaL}), and (\ref{sigma_scaS}), we obtain the absorbing and scattering efficiencies 
as
\begin{subequations}
\begin{eqnarray}
Q_\mathrm{abs}^\mathrm{FOC} &=& -
\frac{4 }{x_\LL ^2}
  \re
   \sum_{n=0}^\infty
  (2n+1) |g_n|^2\left[
   s_{n}^{(\LL)} + \left|s_{n}^{(\LL)}\right|^2
   + n(n+1)\left(\frac{c_\SSS}{c_\LL}\right)^3   \left|s_{n}^{(\SSS,2)}\right|^2
   \right],
   \label{Q_abs_focused}\\
\label{Q_sc_focused}
Q_\mathrm{sca}^\mathrm{FOC}&= &
\frac{4  }{x_\LL^2}
   \sum_{n=0}^\infty
    (2n+1)|g_n|^2
   \left[
       \left|s_{n}^{(\LL)}\right|^2
       +
       n(n+1)\left(\frac{c_\SSS}{c_\LL}\right)^3
           \left|s_{n}^{(\SSS,2)}\right|^2
           \right].
\end{eqnarray}
\end{subequations}
It should be noticed that in the weakly focused regime, $r_0\gg b$, the scattering efficiency is equal to
that of a longitudinal plane wave
\begin{equation}
\label{Q_sc_focused2}
Q_\mathrm{sca}^\mathrm{FOC} = 
 Q_\mathrm{sca}^\LPW.
\end{equation}

The optical theorem for 
Thus,
\begin{eqnarray}
\nonumber
\sigma_\mathrm{ext}^\mathrm{FOC} &=&
   \frac{4 \pi}{k_\LL^2}
   \im\left[
   -\ii
   \sum_{n=0}^\infty
   \left(2n+1\right)|g_n|^2
   s_{n}^{(\LL)}\right]\\
   &=& \frac{4 \pi}{k_\LL^2} \im \left[\vec{f}^{(\LL)}(0,0)\cdot \vec{e}_z\right].
   \label{sigma_ext_focused}  
\end{eqnarray}
Thus, in the on-focus scattering configuration,
the optical theorem has the same format as for a longitudinal plane wave.

We may compute the elastic radiation force exerted on the on-focus sphere 
totally immersed in the focal region.
In such situation, the beam's wavefront can be approached to a traveling plane wave~\cite{Lucas1982}.
Therefore, after inserting Eqs.~(\ref{Q_abs_focused}) and (\ref{Q_sc_focused}) into Eqs.~(\ref{RF}) and (\ref{avg_cos_theta}),
we find the axial radiation force efficiency as
\begin{equation}
Q_{\rad}^\mathrm{FOC} = -
\frac{4 }{x_\LL^2  }      \re
      \sum_{n=0}^\infty \biggl[
      \left(2n+1\right) |g_n|^2
      s_{n}^{(\LL)}
      + 2(n+1) g_{n+1}^*g_n
      \biggl\{
      s_{n+1}^{(\LL)*}s_{n}^{(\LL)}
      +n  (n+2)\left(\frac{c_\SSS}{c_\LL}\right)^3
      s_{n+1}^{(\SSS,2)*}s_{n}^{(\SSS,2)}
      \biggr\} \biggr].
   \label{RF_focus_longitudinal}
\end{equation}
This efficiency has the same structure of that for a longitudinal plane wave.
Though it carries information on diffraction properties of the beam through the $g_n$-coefficients.


\section{Numerical results}
To numerically evaluate the efficiencies which are given by an infinite series, we have to establish a truncation order.
Consider that $Q_n$ is the $n$th-partial term of the efficiencies $Q_\mathrm{sca}$ and $Q_\rad$.
Both efficiency series are truncated at the smallest positive integer $N$ to which the condition $|Q_{N+1}|/|\sum_{n=0}^N Q_n|<10^{-6}$ is satisfied.
Furthermore, the scattering coefficients given in \ref{app:scatt_coefficients} are used to compute $Q_\mathrm{sca}$ and $Q_\rad$.

\subsection{Scattering in aluminium matrix}
The scattering and elastic radiation force efficiencies are computed
for  an iron sphere embedded in an aluminum matrix.
The physical parameters describing these materials are, respectively,
$\rho_1 = \SI{7700}{\kilogram\per \meter\cubed}$, $c_{\LL,1}= \SI{5790}{\meter\per\second}$, $c_{\SSS,1}= \SI{3100}{\meter\per\second}$;
and
$\rho_0 = \SI{2700}{\kilogram\per \meter\cubed}$, $c_\LL= \SI{6568}{\meter\per\second}$, $c_\SSS= \SI{3149}{\meter\per\second}$.
As an initial test, we thoroughly reproduced the results presented in Ref.~\cite{Flax1980}.
Also, we numerically obtained the shear scattering cross-section as given in Ref.~\cite{Korneev1996}.
In both tests, we found excellent agreement with previous results.
For the sake of brevity, we will not show these tests here.

\begin{figure}
\centering
\includegraphics[scale=.5]{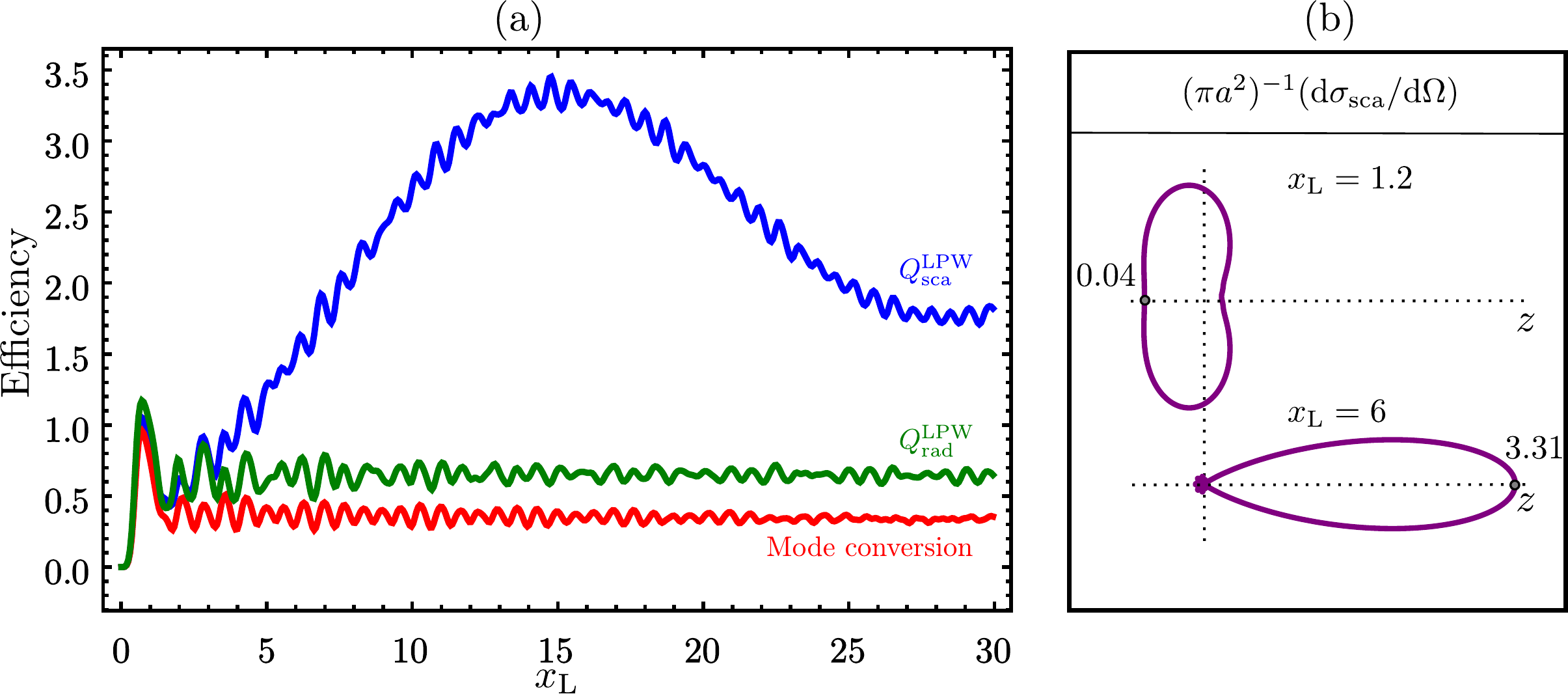} 
\caption{\label{fig:LPW_efficiencies}
(Color online) (a) The scattering $Q_\mathrm{sca}^\LPW$ and radiation force $Q_\rad^\LPW$ efficiencies versus 
the sphere size factor
$x_\LL$ for a longitudinal plane wave (LPW) scattered by an iron sphere embedded in an aluminum matrix.
The mode conversion in scattering is also depicted (red solid line).
(b) The normalized differential scattering-cross section computed 
for $x_\LL=1.2$ and $x_\LL= 6$.
}
\end{figure}

In Fig.~\ref{fig:LPW_efficiencies}.a, we show the scattering and
radiation force efficiencies of a longitudinal plane wave (LPW).
The contribution of mode conversion to $Q_\mathrm{sca}$ is also depicted.
We see that mode conversion is dominant in the band $x_\LL<2$. 
Furthermore, both efficiencies have practically the same magnitude when $x_\LL<3$.
Rapid fluctuations in the efficiencies, due to resonances~\cite{Brill1987}, are observed.
Around $x_\LL=1.2$, we have
$Q_\mathrm{rad}^\LPW<Q_\mathrm{sca}^\LPW$, so according to Eq.~\eqref{ineq_cos_theta}, the asymmetry factor should be negative, $\langle \cos \theta \rangle<0$. 
This is confirmed in Fig.~\ref{fig:LPW_efficiencies}.b by noting that scattering is more prominent in the backward direction.
When $x_\LL>2$, the opposite happens $Q_\mathrm{rad}^\LPW > Q_\mathrm{sca}^\LPW$ and thus forward scattering dominates.
We also observe that the scattering efficiency slowly converges to $2$ as $x_\LL$ increases.
This suggests that longitudinal waves follows the extinction paradox,
which states that a very large sphere blocks twice its cross-sectional area~\cite[p. 68]{Newton1982}.

\begin{figure}
\centering
\includegraphics[scale=.5]{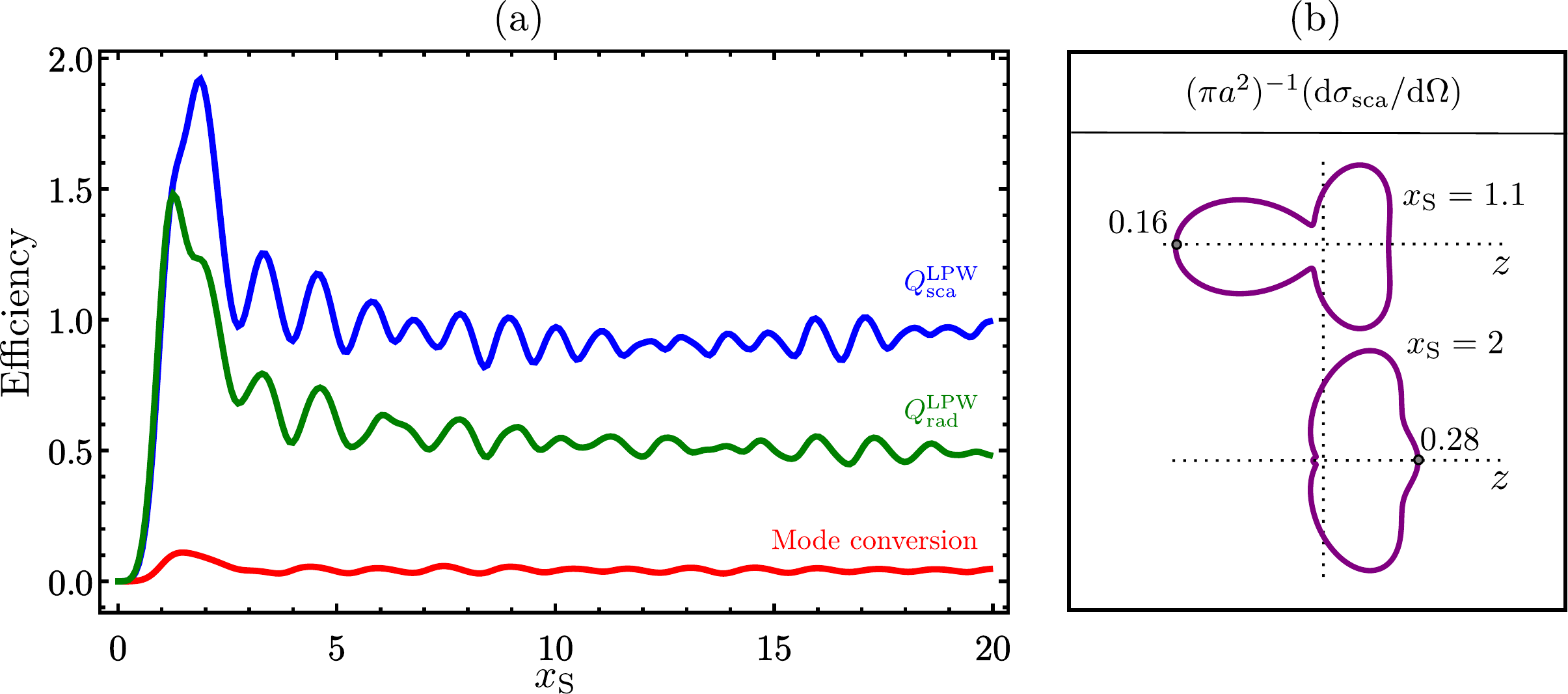} 
\caption{\label{fig:SPW_efficiencies}
(Color online)(Color online) (a) The scattering $Q_\mathrm{sca}^\SPW$ and radiation force $Q_\rad^\SPW$ efficiencies versus 
the sphere size factor
$x_\SSS$ for a shear plane wave (SPW) scattered by an iron sphere embedded in an aluminum matrix.
The mode conversion  in scattering is also shown (red solid line).
(b) The normalized differential scattering-cross section computed 
for $x_\LL=1.1$ and $x_\LL= 2$.}
\end{figure}

The radiation force and scattering efficiencies of a shear plane wave polarized along the $x$-axis 
is plotted in Fig.~\ref{fig:SPW_efficiencies}.a.
The contribution from mode conversion to the scattering efficiency is illustrated.
It is noticed that mode conversion plays a minor role in the current case.
Ripples are observed in both efficiencies due to resonances in the sphere.
In the band $x_\SSS<1.2$ we have
$Q_\mathrm{rad}^\SPW > Q_\mathrm{sca}^\SPW$, and thus the asymmetry factor is negative $\langle \cos \theta \rangle<0$
as discussed in Eq.~\eqref{ineq_cos_theta}.
In Fig.~\ref{fig:SPW_efficiencies}.b, we observe that backscattering is dominant at $x_\SSS=1.1$.
The scattering efficient becomes larger than the radiation force efficiency when $x_\SSS>1.2$.
In this case, forward scattering is dominant and $\langle \cos \theta \rangle>0$ as depicted in Fig.~\ref{fig:SPW_efficiencies}.b.

\begin{figure}
\centering
\includegraphics[scale=.5]{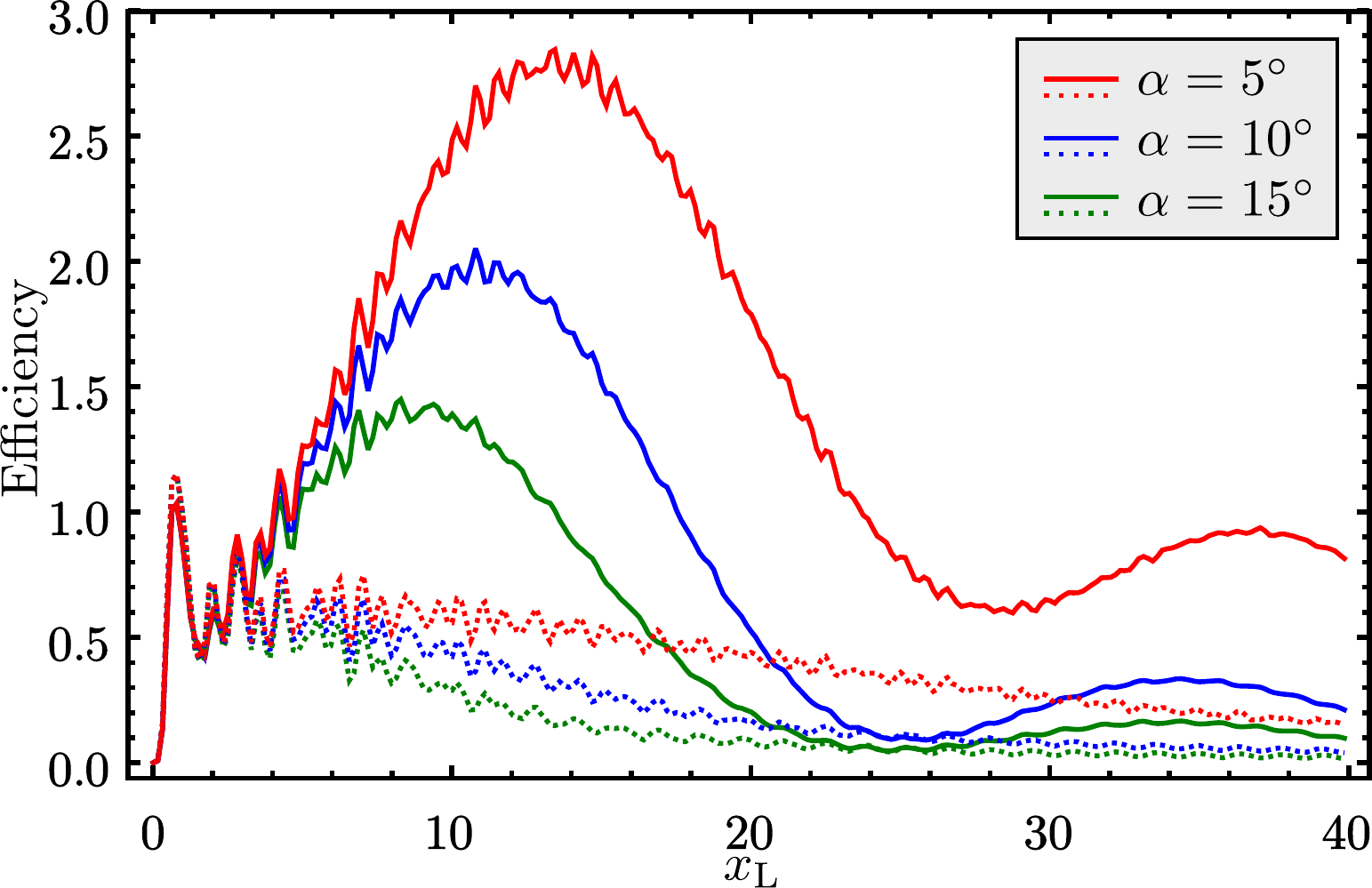} 
\caption{\label{fig:focus_scattering}
(Color online) The scattering  and radiation force  efficiencies (solid and dotted lines, respectively) versus the size factor $x_\LL$ for a longitudinal focused   beam scattered by 
an iron sphere in an aluminum matrix.
The transducer half-aperture angles are $\alpha_0=5^\circ,10^\circ,15^\circ$.
}
\end{figure}

We show the scattering (solid line) and radiation force (dotted line) efficiencies as a function of $x_\LL$ for a longitudinal focused beam in Fig.~\ref{fig:focus_scattering}.
The transducer half-aperture angles are $\alpha_0=5^\circ, 10^\circ, 15^\circ$.
When  $\alpha_0=5^\circ$, the efficiency approaches that of a longitudinal plane wave.
As the beam becomes more focused, the efficiencies decrease.
Also, rapid fluctuations due to resonances in the sphere are observed on both efficiencies.
As the sphere size parameter $x_\LL$ increases, the efficiencies become weaker, i.e. less scattering power is expected.
It should be remembered that to compute the radiation force efficiency given in Eq.~\eqref{RF_focus_longitudinal}, 
we have assumed that the sphere is placed at the transducer focus,
thoroughly inside the focal region.
Such hypothesis is necessary because the energy-momentum relation in Eq.~\eqref{energy-momentum}
is strictly valid within the focal region in which the wavefronts are nearly plane.
Since the beam-waist at $\SI{3}{\decibel}$-intensity is~\cite[p. 185]{Kino1987}
$3.20/(k_\LL \sin \alpha_0)$, it is prudent to use Eq.~\eqref{RF_focus_longitudinal} within the band $x_\LL<3.20/\sin \alpha_0 = 37, 18, 12$ for $\alpha_0=5^\circ, 10^\circ, 15^\circ$, respectively.

\subsection{Elastic radiation force in tissue-like medium}
The importance of elastic radiation force to ultrasound elastography has prompted us to 
analyze this force in tissue-like medium using the developed theory here.
In so doing, we use the descriptive parameters for gel~\cite{Chen2004}:
density $\rho_0 = \SI{1100}{\kilogram \per\meter\cubed}$,
longitudinal speed of sound $c_\LL = \SI{1500}{\meter\per\second}$,
shear elasticity modulus $\mu_{0,\mathrm{E}}=\SI{5360}{\pascal}$,
and shear viscosity $\mu_{0,\mathrm{V}}=\SI{0.36}{\pascal\cdot \second}$.
The shear speed of sound is calculate through the formula~\cite{Yamakoshi1990}
\begin{equation}
c_\SSS =  \sqrt{\frac{2\left(\mu_{0,\mathrm{E}}^2 + \omega^2 \mu_{0,\mathrm{V}}^2 \right)}
{\rho_0 \left(\mu_{0,\mathrm{E}} + \sqrt{\mu_{0,\mathrm{E}}^2 + \omega^2 \mu_{0,\mathrm{V}}^2}\right)}}.
\end{equation}
For a typical ultrasound frequency in elastography $\omega/2\pi =\SI{2.25}{\mega\hertz}$, we have
$c_\SSS=\SI{99.3}{\meter \per \second}$.
The target sphere is assumed to be made of stainless steel (type 4310) with the following parameters:
density $\rho_1=\SI{7840}{\kilogram\per \meter \cubed}$,
longitudinaland shear
speed of sound  $c_{\LL,1}=\SI{5854}{\meter \per \second}$  and $c_{\SSS,1}=\SI{3150}{\meter \per \second}$.
Effects of medium absorption are not considered in this analysis.
We will compared the results in gel to those in water
at room temperature ($\rho_0 = \SI{1000}{\kilogram\per \meter \cubed}$, 
$c_\LL=\SI{1500}{\meter \per \second}$, $c_\SSS=0$).

\begin{figure}
\centering
\includegraphics[scale=.5]{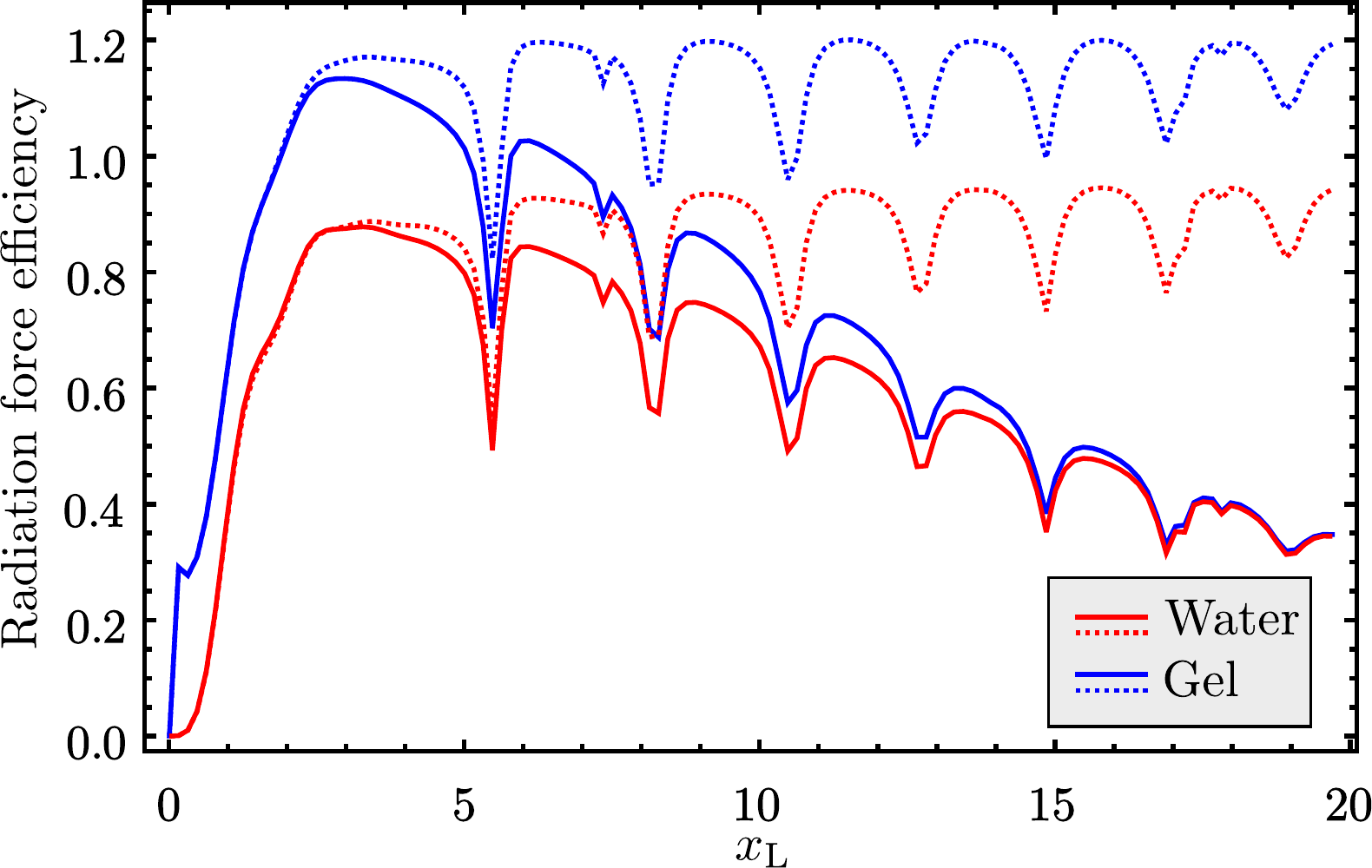} 
\caption{\label{fig:RF_soft}
(Color online) The elastic radiation force efficiency versus the size factor $x_\LL$ for a longitudinal focused  beam (solid lines)
and a longitudinal plane wave (dotted lines) scattered by 
a stainless steel sphere (type 4310) in gel and water.
The ultrasound frequency is $\SI{2.25}{MHz}$ and $\alpha_0=11.45^\circ$.
}
\end{figure}

In Fig.~\ref{fig:RF_soft}, we show the radiation force efficiency for a longitudinal focused beam
with $\alpha_0=11.45^\circ$ and a longitudinal plane wave versus the sphere size parameter $x_\LL$.
The frequency is fixed at $\SI{2.25}{\mega\hertz}$.
It is noticeable that the radiation force of the focused beam in water is in excellent agreement to what
has been obtained by Chen and Apfel~\cite{Chen1996}.
In gel, the shear speed of sound corresponds to $6.6\%$ of its longitudinal counterpart.  
However, a considerable deviation between the efficiencies in gel and water 
is observed.
For both incident waves, the maximum relative difference, adopting the value in gel as the reference, is $98\%$ at $x_\LL=0.16$.
Thus, the prediction from the lossless liquid and soft solid significantly deviates in the long-wavelength limit $x_\LL\ll 1$.
The relative difference at $x_\LL=3.4$ is $23\%$ and still significant.
This value remains the average difference between the plane wave efficiencies.

Some remarks on the radiation force efficiency of the longitudinal focused beam should been drawn.
As the ratio of the medium sound speeds approaches zero, $c_\SSS/c_\LL\rightarrow 0$,
the efficiency in a soft solid is expected to become that in a lossless liquid. 
We observe that in Fig.~\ref{fig:RF_soft} by noting that radiation force efficiency in water 
as computed by Chen and Apfel~\cite{Chen1996} is thoroughly recovered.
Speculatively, this hints that the radiation force efficiency in Eq.~\eqref{RF_focus_longitudinal} is valid for any sphere size 
parameter $x_\LL>0$.
Holding this view, we notice that the difference of the efficiencies in gel and water fades away for $x_\LL>14$. 
Thus, the contribution of mode conversion in this band becomes weaker.

\section{Summary and conclusions}
The extended optical theorem in elastodynamics relates the absorption, scattering, and extinction powers by expressing them in terms of  scattering coefficients. 
On its turn, these coefficients are computed by solving the system of linear equations derived from appropriate continuity conditions of the displacement vectors
and stress fields across the inclusion's surface.
The developed formalism can be applied to the scattering of a longitudinal and shear beam with arbitrary wavefront
by a spherical inclusion made of any material.

We have revisited the classical problem of plane wave scattering and analyzed the contribution of mode conversion to the scattered waves.
The optical theorem for the scattering of a spherically focused beam
by an on-focused sphere was established.
We have derived for the first time the elastic radiation force exerted on a sphere in a solid matrix considering a plane wave and a longitudinal focused beam.
The case of an iron sphere embedded in an aluminum solid matrix was examined.
Additionally, the radiation force on a stainless steel sphere embedded in a gel (soft solid) was computed.
The relative difference between our model and previous water-like medium approach can be as high as $98\%$ in 
the long-wavelength limit.
Therefore, assuming lossless liquid models to estimate the radiation force in soft solids
cannot be taken for granted.

In conclusion, the extended optical theorem can be used 
as a tool to compute the extinction, absorption and scattering powers for arbitrary beams such as Gaussian, Bessel, Airy, etc.
Such analysis may foster new applications in ultrasonic nondestructive testing and geophysics.
Furthermore, the optical theorem provides an elegant and simple way to obtain the elastic radiation force in solids.
We expect that our work will have direct applications in the development and enhancement of elastography methods.

\section*{Acknowledgements}
This work was partially supported by  CNPq (Brazilian agency), Grant No. 303783/2013-3.
 
\appendix

\section{Vector spherical harmonics}
\label{app:VSH}
The vector spherical harmonics (VHS) satisfy the
orthogonality relations~\cite{Barrera1985} 
\begin{subequations}
\label{app:ortho}
\begin{eqnarray}
&& \oint_{4\pi} \vec{Y}_{nm} \cdot \vec{Y}_{n_1m_1}^*
\,
\dd \Omega = \delta_{n,n_1} \delta_{m,m_1},\\
&& \oint_{4\pi} \vec{\Psi}_{nm} \cdot \vec{\Psi}_{n_1m_1}^*
\,
\dd \Omega = \oint_{4\pi} \vec{\Phi}_{nm} \cdot \vec{\Phi}_{n_1m_1}^* 
\,
\dd \Omega = n(n+1) \delta_{n,n_1} \delta_{m,m_1},\\
&& \oint_{4\pi} \vec{Y}_{nm} \cdot \vec{\Psi}_{n_1m_1}^*\dd \Omega
 =
 \oint_{4\pi}\vec{Y}_{nm} \cdot \vec{\Phi}_{n_1m_1}^*\dd \Omega
=  \oint_{4\pi}\vec{\Phi}_{nm} \cdot \vec{\Psi}_{n_1m_1}^*\dd \Omega
= 0.
\end{eqnarray}
\end{subequations}
From VSH definition in (\ref{VSH}), we have 
\begin{subequations}
\label{VHS_psi_phi}
\begin{eqnarray}
\vec{\Psi}_{nm} &=&
(\partial_\theta Y_n^m )\vec{e}_\theta
+
\frac{1}{\sin \theta}(\partial_\varphi Y_n^m) \vec{e}_\varphi,
\\
\vec{\Phi}_{nm}&=&
-\frac{1}{\sin \theta}( \partial_\varphi Y_n^m) \vec{e}_\theta
+(\partial_\theta Y_n^m) \vec{e}_\varphi.
\end{eqnarray} 
\end{subequations}
It follows from Eq.~(\ref{SH}) and the definition of the associate Legendre functions~\cite{Abramowitz1964}
that
$Y_n^m(0,0)=\sqrt{(2n+1)/4\pi} \delta_{m,0}$.
Thus, the VSH 
in the forward scattering direction 
are obtained using this identity and the equations in (\ref{VHS_psi_phi}),
\begin{subequations}
\label{VSH_forward}
\begin{eqnarray}
\vec{Y}_{nm}(0,0)&=& \sqrt{\frac{2n +1}{4 \pi}} \delta_{m,0} \, \vec{e}_z,\\
 \vec{\Psi}_{nm}(0,0)&=&
  \frac{1}{2} \sqrt{\frac{2n+1}{4 \pi}}
 \sqrt{n(n+1)}
 [(\delta_{m,-1}-\delta_{m,1})\vec{e}_x
 - \ii (\delta_{m,-1} + \delta_{m,1})\vec{e}_y],
 \\ 
 \vec{\Phi}_{nm}(0,0)&=&
 \frac{\ii}{2} \sqrt{\frac{2n+1}{4 \pi}}
  \sqrt{n(n+1)}
  [(\delta_{m,-1}+\delta_{m,1})\vec{e}_x - \ii (\delta_{m,-1} - \delta_{m,1})\vec{e}_y],
\end{eqnarray}
\end{subequations}
where $\vec{e}_x$, $\vec{e}_y$, and $\vec{e}_z$
are the Cartesian unit-vectors.
Similarly, using the expression $Y_n^m(\pi,0)= (- 1)^nY_n^m(0,0)$, we find the VSHs in the backscattering direction $\theta=\pi$,
\begin{subequations}
\label{VSH_backward}
\begin{eqnarray}
\vec{Y}_{nm}(\pi,0)&=&  (-1)^{n+1}\vec{Y}_{nm}(0,0),\\
 \vec{\Psi}_{nm}(\pi,0)&=& (-1)^{n+1}\vec{\Psi}_{nm}(0,0),\\
 \vec{\Phi}_{nm}(\pi,0)&=& (-1)^{n+1}\vec{\Phi}_{nm}(0,0).
\end{eqnarray}
\end{subequations}
We have used the relations $ \vec{e}_r(0,0) = \vec{e}_z$,
$\vec{e}_r(\pi,0)= - \vec{e}_z$,
$\vec{e}_\theta(0,0) = \vec{e}_x$, $\vec{e}_\theta(\pi,0) = -\vec{e}_x$, 
$\vec{e}_\varphi(\pi,0) = \vec{e}_\varphi(0,0)=\vec{e}_y$.

Other common definition of VSH is~\cite[p.~1899]{Morse1953a}
\begin{subequations}
\begin{eqnarray}
\vec{P}_{nm} &=&
(-1)^m
 \sqrt{\frac{4 \pi (n + m)!}{(2n + 1 ) (n - m)!}}
 \vec{Y}_{nm},\\
 \vec{B}_{nm} &=&
 \frac{(-1)^m}{\sqrt{n(n+1)}}
  \sqrt{\frac{4 \pi (n + m)!}{(2n + 1 ) (n - m)!}}
  \vec{\Psi}_{nm},\\
  \vec{C}_{nm} &=&
   \frac{(-1)^{m+1}}{\sqrt{n(n+1)}}
    \sqrt{\frac{4 \pi (n + m)!}{(2n + 1 ) (n - m)!}}
    \vec{\Phi}_{nm}.
\end{eqnarray} 
\end{subequations}
Using the recursion expressions for these vectors given in~\cite{Clapp1970}, we find 
\begin{subequations}
\label{app:recurrence}
\begin{eqnarray}
\cos \theta\, \vec{Y}_{nm}
&=&
\sqrt{\frac{(n-m)(n+m)}{(2n-1)(2n+1)}}\vec{Y}_{n-1,m}
+
\sqrt{\frac{(n-m+1)(n+m+1)}{(2n+1)(2n+3)}}\vec{Y}_{n+1,m},
 \\
 \nonumber
\cos \theta \, \vec{\Psi}_{nm}
&= &\frac{n+1}{n}
\sqrt{\frac{(n-m)(n+m)}{(2n-1)(2n+1)}}\vec{\Psi}_{n-1,m}
+\frac{\ii m}{n(n+1)}\vec{\Phi}_{nm}
\\
&+&
\frac{n}{n+1}\sqrt{\frac{(n-m+1)(n+m+1)}{(2n+1)(2n+3)}}\vec{\Psi}_{n+1,m},\\
\nonumber
\cos \theta \, \vec{\Phi}_{nm}
&=&
\frac{n+1}{n}
\sqrt{\frac{(n-m)(n+m)}{(2n-1)(2n+1)}}\vec{\Phi}_{n-1,m}
- \frac{\ii m}{n(n+1)}\vec{\Psi}_{nm}
\\
&+&
\frac{n}{n+1}\sqrt{\frac{(n-m+1)(n+m+1)}{(2n+1)(2n+3)}}\vec{\Phi}_{n+1,m}.
\end{eqnarray}
\end{subequations}


\section{Longitudinal and shear absorption power components}

Using Eqs.~(\ref{uinL_far}) and (\ref{uinMN_far}),
we find that the terms of the absorbing power 
given in Eq.~(\ref{abs3}) at the farfield are
\begin{subequations}
\label{power_components}
\begin{eqnarray}
\re\left[ \ii u_{r,\mathrm{in}}^{(\LL)}\partial_r u_{r,\mathrm{sc}}^{(\LL)\,*} 
  + \ii  u_{r,\mathrm{sc}}^{(\LL)}\partial_r u_{r,\mathrm{in}}^{(\LL)\,*} \right] &=&
   \frac{1}{k_\LL r^2}\re \sum_{\substack{n,m\\n_1,m_1}}     
    s_{nm}^{(\LL)} a_{n_1m_1}^{(\LL)\,*}
  Y_n^m Y_{n_1}^{m_1\,*},\\
 \re \left[  \ii u_{r,\mathrm{sc}}^{(\LL)}\partial_r u_{r,\mathrm{sc}}^{(\LL)\,*}\right] &=&
  \frac{1}{k_\LL r^2}\re \sum_{\substack{n,m\\n_1,m_1}}
  s_{nm}^{(\LL)} s_{n_1m_1}^{(\LL)\,*} 
  Y_n^m Y_{n_1}^{m_1\,*},\\
\nonumber
\re \left[\ii \vec{u}_{\mathrm{sc}}^{(\SSS)}\cdot \partial_r \vec{u}_{\mathrm{in}}^{(\SSS)\,*}
+ \ii \vec{u}_{\mathrm{in}}^{(\SSS)}\cdot \partial_r \vec{u}_{\mathrm{sc}}^{(\SSS)\,*}\right] &=& 
\frac{1}{k_\SSS r^2}\re \sum_{\substack{n,m\\n_1,m_1}} 
\bigl[
s_{nm}^{(\SSS,1)} a_{n_1m_1}^{(\SSS,1)\,*}
\vec{\Phi}_{nm}\cdot \vec{\Phi}_{n_1m_1}^*
+
s_{nm}^{(\SSS,2)} a_{n_1m_1}^{(\SSS,2)\,*}
\vec{\Psi}_{nm}\cdot \vec{\Psi}_{n_1m_1}^*
\bigr], \\ \\
\nonumber
\re \left[
\ii \vec{u}_{\mathrm{sc}}^{(\SSS)}\cdot \partial_r \vec{u}_{\mathrm{sc}}^{(\SSS)\,*}\right] &=&
\frac{1}{k_\SSS r^2}\re \sum_{\substack{n,m\\n_1,m_1}} 
\bigl[ 
s_{n,m}^{(\SSS,1)} s_{n_1,m_1}^{(\SSS,1)\,*}
\vec{\Phi}_{nm}\cdot \vec{\Phi}_{n_1m_1}^*
+ s_{n,m}^{(\SSS,2)} s_{n_1,m_1}^{(\SSS,2)\,*}\vec{\Psi}_{nm} \cdot \vec{\Psi}_{n_1m_1}^*\bigr].\\
\end{eqnarray}
\end{subequations}


\section{Scattering coefficients}
\label{app:scatt_coefficients}
To calculate the longitudinal scattering coefficients we proceed as follows.
The incident, scattered, and transmitted displacement vectors are given, respectively, in Eqs.~\eqref{set_solution}, \eqref{uscattered}, and \eqref{u_transmitted}.
The corresponding stress tensors are calculated by inserting the displacement vectors into Eq.~\eqref{stress-strain}.
The obtained displacements and stresses are then substituted in the boundary conditions~\eqref{boundary_L}.
Hence, we find a system of linear equations for the unknown coefficients,
\begin{equation}
\mathbf{D}_n \vec{x}_n^T=
\left[
\begin{matrix}
d_{11} & d_{12} & d_{13} & d_{14}\\
d_{21} & d_{22} & d_{23} & d_{24}\\
d_{31} & d_{32} & d_{33} & d_{34} \\
d_{41} & d_{42} & d_{43} & d_{44}\\
\end{matrix}
\right]
\left[
\begin{matrix}
s_n^{(\LL)}\\ s_n^{(\SSS)}\\ t_n^{(\LL)}\\ t_n^{(\SSS)}
\end{matrix}
\right] =
\left[
\begin{matrix}
b_1 \\ b_2 \\ b_3\\ b_4
\end{matrix}
\right] = \vec{b}_n^T. 
\label{Sn}
\end{equation}
According to the Cramer's rule, the scattering coefficients
are given by
\begin{equation}
\label{scaled_scatt_coeff}
s_n^{(\LL)} = \frac{\det \mathbf{D}_n^{(\LL)}}{\det \mathbf{D}_n}, \quad
s_n^{(\SSS)} =\frac{c_\LL}{c_\SSS}\frac{\det \mathbf{D}_n^{(\SSS)}}{\det \mathbf{D}_n},
\end{equation}
where 
the matrices
$\mathbf{D}_n^{(\LL)}$ and
$\mathbf{D}_n^{(\SSS)}$ are given in terms of $\mathbf{D}_n$, except by replacing, respectively, its first and second columns
by $\vec{b}_n^T$.
The elements of $\mathbf{D}_n$ are given by 
\begin{eqnarray}
\nonumber
&& d_{11} =  x_\LL \, h_n^{(1)\prime}(x_\LL),\\
\nonumber
&& d_{12} = n(n+1) h_n^{(1)}(x_\SSS), \\
\nonumber
&& d_{13} = - x_{\LL,1}\, j_n'(x_{\LL,1}),\\ 
\nonumber
&& d_{14} = - n(n+1) j_n(x_{\SSS,1}),\\
\nonumber
&& d_{21} = h_n^{(1)}( x_\LL),\\
\nonumber
&& d_{22} = x_\SSS \, h_n^{(1)\prime}(x_{\SSS})+  h_n^{(1)}(x_{\SSS}), \\
\nonumber
&& d_{23} = - j_n( x_{\LL,1}),\\
\nonumber
&& d_{24} = 
- \left[x_{\SSS,1} \, j_n'( x_{\SSS,1})+ j_n( x_{\SSS,1})  \right]
,\\
\nonumber
&& d_{31} =  
 [2 n(n+1) - x_\SSS^2] 
h_n^{(1)}(x_\LL) - 4   x_\LL\,
h_n^{(1)\prime}(x_\LL),\\
\nonumber
&& d_{32} =
2 n(n+1)
\left[ x_\SSS \,
h_n^{(1)\prime}(x_\SSS) - h_n^{(1)}(x_{\SSS})
\right],\\
\nonumber
&& d_{33} = -
\frac{\rho_1}{\rho_0}  \left(\frac{c_{\SSS,1}}{c_\SSS}\right)^2
\bigl[[2 n(n+1) - x_{\SSS,1}^2 ]
j_n(x_{\LL,1}) - 4 x_{\LL,1}\, j_n'(x_{\LL,1})
\bigr],\\
\nonumber
&& d_{34} = - 2 \frac{\rho_1 }{\rho_0}\left(\frac{c_{\SSS,1}}{c_\SSS}\right)^2 n(n+1)
\bigl[ x_{\SSS,1} \, j_n'(x_{\SSS,1})-j_n(x_{\SSS,1})
\bigr],\\
&& d_{41} =
x_\LL\,
h_n^{(1)\prime}(x_\LL) - h_n^{(1)}(x_\LL),\\
\nonumber
&& d_{42} =
\biggl[
\left(n(n+1) - \frac{x_\SSS^2}{2} - 1  \right) h_n^{(1)}(x_\SSS) -  x_\SSS\, h_n^{(1)\prime}(x_\SSS)
\biggr],\\
\nonumber
&& d_{43} = -
\frac{\rho_1 }{\rho_0}\left(\frac{c_{\SSS,1}}{c_\SSS}\right)^2
\left[  x_{\LL,1}\,  j_n'(x_{\LL,1}) - j_n(x_{\LL,1})
\right],
\\
\nonumber
&& d_{44} = -  \frac{\rho_1 }{\rho_0}\left(\frac{c_{\SSS,1}}{c_\SSS}\right)^2
\biggl[
\left( n(n+1) - \frac{x_{\SSS,1}^2}{2} -1
\right) j_n(x_{\SSS,1}) -
 x_{\SSS,1} \, j_n'(x_{\SSS,1})
\biggr],\\
\nonumber
&& b_1 = -x_\LL j_n'(x_\LL),\\
\nonumber
&& b_2 = - j_n(x_\LL),\\
\nonumber
&& b_3 = - 
[2 n(n+1) - x_\SSS^2] 
j_n(x_\LL) + 4   x_\LL\,
j_n'(x_\LL),\\
\nonumber
&& b_4 =-
x_\LL\,
j_n'(x_\LL) + j_n(x_\LL).
\end{eqnarray}
These elements have also been obtained in Ref.~\cite{Brill1987}.

Likewise the longitudinal case, the shear scattering coefficients are obtained from the linear system of equations
\begin{equation}
\mathbf{D}_n \vec{x}_n^T=
\left[
\begin{matrix}
d_{11} & d_{12} & d_{13} & d_{14}& 0 & 0\\
d_{21} & d_{22} & d_{23} & d_{24}& 0 & 0\\
d_{31} & d_{32} & d_{33} & d_{34}& 0 & 0 \\
d_{41} & d_{42} & d_{43} & d_{44}& 0 & 0\\
0 & 0 & 0 & 0& d_{55} & d_{56} \\
0 & 0 & 0 & 0& d_{65} & d_{66}\\
\end{matrix}
\right]
\left[
\begin{matrix}
s_n^{(\LL)}\\ s_n^{(\SSS,2)}\\ t_n^{(\LL)}\\ t_n^{(\SSS,2)}\\s_n^{(\SSS,1)}\\t_n^{(\SSS,1)}
\end{matrix}
\right] =
\left[
\begin{matrix}
b_1 \\ b_2 \\ b_3\\ b_4 \\ b_5 \\ b_6
\end{matrix}
\right] = \vec{b}_n^T. 
\label{SS}
\end{equation}
According to the Cramer's rule, the scattering coefficients
are given by
\begin{equation}
\label{scaled_scatt_coeff2}
s_n^{(\LL)} = \frac{c_\SSS}{c_\LL}\frac{\det \mathbf{D}_n^{(\LL)}}{\det \mathbf{D}_n}, \quad
s_n^{(\SSS,2)} =\frac{\det \mathbf{D}_n^{(\SSS,2)}}{\det \mathbf{D}_n}, \quad
s_n^{(\SSS,1)} =\frac{\det \mathbf{D}_n^{(\SSS,1)}}{\det \mathbf{D}_n}.
\end{equation}
Here the matrices
$\mathbf{D}_n^{(\LL)}$, $\mathbf{D}_n^{(\SSS,1)}$, and $\mathbf{D}_n^{(\SSS,2)}$
are obtained from $\mathbf{D}_n$ by, respectively, replacing its first, second, and third columns by 
$\vec{b}_n^T$.
The additional matrix  elements required to compute the scattering coefficients are 
\begin{eqnarray}
\nonumber
&& d_{55} =  x_\SSS\, h_n^{(1)}(x_{\SSS}),\\
\nonumber
&& d_{56} = - x_\SSS\,j_n(x_{\SSS,1}) \\
\nonumber
&& d_{65} =  x_\SSS\,h_n^{(1)\prime}(x_\SSS)-h_n^{(1)}(x_\SSS),\\ 
\nonumber
&& d_{66} = - \frac{\rho_1 }{\rho_0}\left(\frac{c_{\SSS,1}}{c_\SSS}\right)
\biggl[x_{\SSS,1} j_n'(x_{\SSS,1})- j_n(x_{\SSS,1})
\biggr],\\
&& b_1 = -n(n+1) j_n(x_\SSS),\\
\nonumber
&& b_2 = - [x_\SSS\,j_n'(x_{\SSS})+j_n(x_{\SSS})],\\
\nonumber
&& b_3 =-2 n(n+1)
\left[ x_\SSS \,
j_n'(x_\SSS)- j_n(x_{\SSS})
\right],\\
\nonumber
&& b_4 =-\biggl[
\left(n(n+1)-\frac{x_\SSS^2}{2} - 1  \right) j_n(x_\SSS) -  x_\SSS\, j_n'(x_\SSS)
\biggr],\\
\nonumber
&& b_5 =-x_\SSS\, j_n(x_\SSS),\\
&& b_6 =-[x_\SSS\,j_n'(x_\SSS)-j_n(x_\SSS)].
\nonumber
\end{eqnarray}


\begin{thebibliography}{10}
\expandafter\ifx\csname url\endcsname\relax
  \def\url#1{\texttt{#1}}\fi
\expandafter\ifx\csname urlprefix\endcsname\relax\def\urlprefix{URL }\fi
\expandafter\ifx\csname href\endcsname\relax
  \def\href#1#2{#2} \def\path#1{#1}\fi

\bibitem{Newton1976}
R.~G. Newton, {Optical theorem and beyond}, Am. J. Phys. 44~(7) (1976)
  639--642.

\bibitem{Mie1908}
G.~Mie, {Beitr{\"{a}}ge zur Optik tr{\"{u}}ber Medien, speziell kolloidaler
  Metall{\"{o}}sungen}, Ann. Phys. 330~(3) (1908) 377--445, in German.
\newblock \href {http://dx.doi.org/10.1002/andp.19083300302}
  {\path{doi:10.1002/andp.19083300302}}.

\bibitem{Feenberg1932}
E.~Feenberg, The scattering of slow electrons by neutral atoms, Phys. Rev.
  40~(1) (1932) 40--54.
\newblock \href {http://dx.doi.org/10.1103/PhysRev.40.40}
  {\path{doi:10.1103/PhysRev.40.40}}.

\bibitem{VanDeHulst1949}
H.~C. van~de Hulst, {On the attenuation of plane waves by obstacles of
  arbitrary size and form}, Physica 15~(8-9) (1949) 740--746.
\newblock \href {http://dx.doi.org/10.1016/0031-8914(49)90079-8}
  {\path{doi:10.1016/0031-8914(49)90079-8}}.

\bibitem{DeHoop1959}
A.~T. {de Hoop}, {On the plane-wave extinction cross-section of an obstacle},
  Appl. Sci. Res. Sect. B 7~(1) (1959) 463--469.
\newblock \href {http://dx.doi.org/10.1007/BF02921932}
  {\path{doi:10.1007/BF02921932}}.

\bibitem{Barratt1965}
P.~J. Barratt, W.~D. Collins, {The scattering cross-section of an obstacle in
  an elastic solid for plane harmonic waves}, Math. Proc. Cambridge Philos.
  Soc. 61~(04) (1965) 969--981.
\newblock \href {http://dx.doi.org/10.1017/S0305004100039360}
  {\path{doi:10.1017/S0305004100039360}}.

\bibitem{Gubernatis1977}
J.~E. Gubernatis, E.~Domany, J.~A. Krumhansl, {Formal aspects of the theory of
  the scattering of ultrasound by flaws in elastic materials}, J. Appl. Phys.
  48~(7) (1977) 2804--2811.
\newblock \href {http://dx.doi.org/10.1063/1.324141}
  {\path{doi:10.1063/1.324141}}.

\bibitem{Varatharajulu1977}
V.~Varatharajulu, {Reciprocity relations and forward amplitude theorems for
  elastic waves}, J. Math. Phys. 18~(4) (1977) 537.
\newblock \href {http://dx.doi.org/10.1063/1.523335}
  {\path{doi:10.1063/1.523335}}.

\bibitem{Korneev1993}
V.~A. Korneev, L.~R. Johnson, Scattering of elastic waves by a spherical
  inclusion--{I. Theory and numerical results}, Geophys. J. Int. 115 (1993)
  230--250.

\bibitem{Korneev1996}
V.~A. Korneev, L.~R. Johnson, {Scattering of P and S waves by a spherically
  symmetric inclusion}, Pure Appl. Geophys. 147 (1996) 675--718.

\bibitem{Glauber1953}
R.~Glauber, V.~Schomaker, The theory of electron diffraction, Phys. Rev. 89~(4)
  (1953) 667--671.
\newblock \href {http://dx.doi.org/10.1103/PhysRev.89.667}
  {\path{doi:10.1103/PhysRev.89.667}}.

\bibitem{Dassios1980}
G.~Dassios, Second order low-frequency scattering by the soft ellipsoid, SIAM
  J. Appl. Math. 38~(3) (1980) 373--381.
\newblock \href {http://dx.doi.org/10.1137/0138031}
  {\path{doi:10.1137/0138031}}.

\bibitem{Marston2001}
P.~L. Marston, {Generalized optical theorem for scatterers having inversion
  symmetry: Applications to acoustic backscattering}, J. Acoust. Soc. Am.
  109~(4) (2001) 1291--1295.
\newblock \href {http://dx.doi.org/10.1121/1.1352082}
  {\path{doi:10.1121/1.1352082}}.

\bibitem{Ratilal2001}
P.~Ratilal, N.~C. Makris, {Extinction theorem for object scattering in a
  stratified medium}, J. Acoust. Soc. Am. 110~(6) (2001) 2924.
\newblock \href {http://dx.doi.org/10.1121/1.1405522}
  {\path{doi:10.1121/1.1405522}}.

\bibitem{Kriegsmann1985}
G.~Kriegsmann, A.~Norris, E.~Reiss, {An “optical” theorem for acoustic
  scattering by baffled flexible surfaces}, J. Sound Vib. 99~(3) (1985)
  301--307.
\newblock \href {http://dx.doi.org/10.1016/0022-460X(85)90369-4}
  {\path{doi:10.1016/0022-460X(85)90369-4}}.

\bibitem{Halliday2009}
D.~Halliday, A.~Curtis, {Generalized optical theorem for surface waves and
  layered media}, Phys. Rev. E 79~(5) (2009) 056603.
\newblock \href {http://dx.doi.org/10.1103/PhysRevE.79.056603}
  {\path{doi:10.1103/PhysRevE.79.056603}}.

\bibitem{Markel1991}
V.~A. Markel, L.~S. Muratov, M.~I. Stockman, T.~F. George, {Theory and
  numerical simulation of optical properties of fractal clusters}, Phys. Rev. B
  43~(10) (1991) 8183--8195.
\newblock \href {http://dx.doi.org/10.1103/PhysRevB.43.8183}
  {\path{doi:10.1103/PhysRevB.43.8183}}.

\bibitem{Lock1995}
J.~A. Lock, J.~T. Hodges, G.~Gouesbet, {Failure of the optical theorem for
  Gaussian-beam scattering by a spherical particle}, J. Opt. Soc. Am. A 12~(12)
  (1995) 2708.
\newblock \href {http://dx.doi.org/10.1364/JOSAA.12.002708}
  {\path{doi:10.1364/JOSAA.12.002708}}.

\bibitem{Gouesbet2009}
G.~Gouesbet, {On the optical theorem and non-plane-wave scattering in quantum
  mechanics}, J. Math. Phys. 50~(11) (2009) 112302.
\newblock \href {http://dx.doi.org/10.1063/1.3256127}
  {\path{doi:10.1063/1.3256127}}.

\bibitem{Zhang2013}
L.~Zhang, P.~L. Marston, Optical theorem for acoustic non-diffracting beams and
  application to radiation force and torque, Biomed. Opt. Express 4~(9) (2013)
  1610--1617.

\bibitem{Mitri2014}
F.~G. Mitri, G.~T. Silva, {Generalization of the extended optical theorem for
  scalar arbitrary-shape acoustical beams in spherical coordinates}, Phys. Rev.
  E 90~(5) (2014) 053204.
\newblock \href {http://dx.doi.org/10.1103/PhysRevE.90.053204}
  {\path{doi:10.1103/PhysRevE.90.053204}}.

\bibitem{Gouesbet2007}
G.~Gouesbet, {Asymptotic quantum inelastic generalized Lorenz–Mie theory},
  Opt. Commun. 278~(1) (2007) 215--220.
\newblock \href {http://dx.doi.org/10.1016/j.optcom.2007.06.006}
  {\path{doi:10.1016/j.optcom.2007.06.006}}.

\bibitem{Mitri2015}
F.~G. Mitri, {Optical theorem for two-dimensional ({2D}) scalar monochromatic
  acoustical beams in cylindrical coordinates.}, Ultrasonics 62 (2015) 20--26.
\newblock \href {http://dx.doi.org/10.1016/j.ultras.2015.02.019}
  {\path{doi:10.1016/j.ultras.2015.02.019}}.

\bibitem{Mitri2016}
F.~G. Mitri, {Extended optical theorem for scalar monochromatic acoustical
  beams of arbitrary wavefront in cylindrical coordinates.}, Ultrasonics 67
  (2016) 129--135.
\newblock \href {http://dx.doi.org/10.1016/j.ultras.2016.01.006}
  {\path{doi:10.1016/j.ultras.2016.01.006}}.

\bibitem{Mitri2015a}
F.~G. Mitri, {Generalization of the optical theorem for monochromatic
  electromagnetic beams of arbitrary wavefront in cylindrical coordinates}, J.
  Quant. Spectrosc. Radiat. Transf. 166 (2015) 81--92.
\newblock \href {http://dx.doi.org/10.1016/j.jqsrt.2015.07.016}
  {\path{doi:10.1016/j.jqsrt.2015.07.016}}.

\bibitem{DeL.Kronig1926}
R.~{de L. Kronig}, On the theory of dispersion of x-rays, J. Opt. Soc. Am.
  12~(6) (1926) 547--557.
\newblock \href {http://dx.doi.org/10.1364/JOSA.12.000547}
  {\path{doi:10.1364/JOSA.12.000547}}.

\bibitem{Toll1956}
J.~S. Toll, Causality and the dispersion relation: Logical foundations, Phys.
  Rev. 104~(6) (1956) 1760--1770.
\newblock \href {http://dx.doi.org/10.1103/PhysRev.104.1760}
  {\path{doi:10.1103/PhysRev.104.1760}}.

\bibitem{Newton1968}
R.~G. Newton, Determination of the amplitude from the differential cross
  section by unitarity, J. Math. Phys. 9~(12) (1968) 2050--2055.
\newblock \href {http://dx.doi.org/10.1063/1.1664543}
  {\path{doi:10.1063/1.1664543}}.

\bibitem{Kraut1976}
E.~Kraut, Review of theories of scattering of elastic waves by cracks, IEEE
  Trans. Sonics Ultrason. 23~(3) (1976) 162--167.
\newblock \href {http://dx.doi.org/10.1109/T-SU.1976.30856}
  {\path{doi:10.1109/T-SU.1976.30856}}.

\bibitem{Kitahara1997}
{M. Kitahara and K. Nakagawa}, Elastodynamic optical theorem for the evaluation
  of scattering cross-sections for a crack, in: D.~O. Thompson, D.~E. Chimenti
  (Eds.), Rev. Prog. Quant. Nondestruct. Eval., Springer US, Boston, MA, 1997,
  pp. 27--34.
\newblock \href {http://dx.doi.org/10.1007/978-1-4615-5947-4}
  {\path{doi:10.1007/978-1-4615-5947-4}}.

\bibitem{Carney1999}
P.~S. Carney, E.~Wolf, G.~S. Agarwal, {Diffraction tomography using power
  extinction measurements}, J. Opt. Soc. Am. A 16~(11) (1999) 2643--2648.

\bibitem{Groenenboom1995}
J.~Groenenboom, R.~Snieder, {Attenuation, dispersion, and anisotropy by
  multiple scattering of transmitted waves through distributions of
  scatterers}, J. Acoust. Soc. Am. 98~(6) (1995) 3482.
\newblock \href {http://dx.doi.org/10.1121/1.413780}
  {\path{doi:10.1121/1.413780}}.

\bibitem{Margerin2011}
L.~Margerin, H.~Sato, {Generalized optical theorems for the reconstruction of
  Green's function of an inhomogeneous elastic medium.}, J. Acoust. Soc. Am.
  130~(6) (2011) 3674--3690.
\newblock \href {http://dx.doi.org/10.1121/1.3652856}
  {\path{doi:10.1121/1.3652856}}.

\bibitem{Wapenaar2010}
K.~Wapenaar, E.~Slob, R.~Snieder, {On seismic interferometry, the generalized
  optical theorem, and the scattering matrix of a point scatterer}, Geophysics
  75~(3) (2010) SA27--SA35.

\bibitem{Maurel2008}
A.~Maurel, V.~Pagneux, F.~Barra, F.~Lund, Interaction between an elastic wave
  and a single pinned dislocation, Phys. Rev. B 72 (2008) 174110.

\bibitem{Krautkramer1990}
J.~Krautkramer, H.~Krautkramer, {Ultrasonic Testing of Materials},
  Springer-Verlag, Berlin, Germany, 1990.

\bibitem{Smyshlyaev1994}
V.~P. Smyshlyaev, J.~R. Willis, Linear and nonlinear scattering of elastic
  waves by microcracks, J. Mech. Phys. Solids 42~(4) (1994) 585--610.

\bibitem{Hasegawa1969}
T.~Hasegawa, K.~Yosioka, Acoustic‐radiation force on a solid elastic sphere,
  J. Acoust. Soc. Am. 46 (1969) 1139--1143.

\bibitem{Chen1996}
X.~Chen, R.~E. Apfel, Radiation force on a spherical object in an axisymmetric
  wave field and its application to the calibration of high-frequency
  transducers, J. Acoust. Soc. Am. 99 (1996) 713--724.

\bibitem{Marston2006}
P.~L. Marston, Axial radiation force of a bessel beam on a sphere and direction
  reversal of the force, J. Acoust. Soc. Am. 120 (2006) 3518--3524.

\bibitem{Mitri2009}
F.~G. Mitri, Negative axial radiation force on a fluid and elastic spheres
  illuminated by a high-order {Bessel} beam of progressive waves, J. Phys. A 42
  (2009) 245202.

\bibitem{Silva2011a}
G.~T. Silva, An expression for the radiation force exerted by an acoustic beam
  with arbitrary wavefront {(L)}, J. Acoust. Soc. Am. 130 (2011) 3541--3544.

\bibitem{Azarpeyvand2012}
M.~Azarpeyvand, Acoustic radiation force of a bessel beam on a porous sphere,
  J. Acoust. Soc. Am. 131 (2012) 4337--4348.

\bibitem{Sapozhnikov2013}
O.~A. Sapozhnikov, M.~R. Bailey, Radiation force of an arbitrary acoustic beam
  on an elastic sphere in a fluid, J. Acoust. Soc. Am. 133 (2013) 661--676.

\bibitem{Baresch2013}
D.~Baresch, J.~L. Thomas, R.~Marchiano, Three-dimensional acoustic radiation
  force on an arbitrarily located elastic sphere., J. Acoust. Soc. Am. 133
  (2013) 25--36.

\bibitem{Wells2011}
P.~N.~T. Wells, H.-D. Liang, Medical ultrasound: imaging of soft tissue strain
  and elasticity, J. R. Soc. Interface\href
  {http://dx.doi.org/10.1098/rsif.2011.0054}
  {\path{doi:10.1098/rsif.2011.0054}}.

\bibitem{Sarvazyan2011}
A.~Sarvazyan, T.~J. Hall, M.~W. Urban, M.~Fatemi, S.~R. Aglyamov, B.~S. Garra,
  An overview of elastography -- an emergin branch of medical imaging, Curr.
  Med. Imaging Rev. 7 (2011) 255--282.

\bibitem{Palmeri2011}
M.~L. Palmeri, K.~R. Nightingale, Acoustic radiation force-based elasticity
  imaging methods, Interface Focus 6 (2011) 553--564.

\bibitem{Bercoff2004}
J.~Bercoff, M.~Tanter, M.~Fink, Supersonic shear imaging: a new technique for
  soft tissue elasticity mapping, IEEE Trans. Ultrason. Ferroelectr. Freq.
  Contr. 51 (2004) 396--409.

\bibitem{Ordeig2016}
O.~Ordeig, S.~Y. Chin, S.~Kim, P.~V. Chitnis, S.~K. Sia, An implantable
  compound-releasing capsule triggered on demand by ultrasound, Sci. Rep.
  6~(22803) (2016) 1--11.

\bibitem{Aglyamov2007}
S.~R. Aglyamov, A.~B. Karpiouk, Y.~A. Ilinskii, E.~A. Zabolotskaya, S.~Y.
  Emelianov, Motion of a solid sphere in a viscoelastic medium in response to
  applied acoustic radiation force: Theoretical analysis and experimental
  verification, J. Acoust. Soc. Am. 122 (2007) 1927--1936.

\bibitem{Andreev2016}
V.~G. Andreev, I.~Y. Demin, Z.~A. Korolkov, A.~V. Shanin, Motion of spherical
  microparticles in a viscoelastic medium under the action of acoustic
  radiation force, Bull. Russ. Acad. Sci. Phys. 80 (2016) 1191--1196.

\bibitem{Fung2001}
Y.~C. Fung, P.~Tong, {Classical and Computational Solid Mechanics}, World
  Scientific, Singapore, 2001.

\bibitem{Morse1953}
P.~M. Morse, H.~Feshbach, {Methods of Theoretical Physics, Part I}, McGraw-Hill
  Inc., New York, NY USA, 1953.

\bibitem{Morse1953a}
P.~M. Morse, H.~Feshbach, {Methods of Theoretical Physics, Part II},
  McGraw-Hill, Inc., New York, NY USA, 1953.

\bibitem{Barrera1985}
R.~G. Barrera, G.~A. Estevez, J.~Giraldo, {Vector spherical harmonics and their
  application to magnetostatics}, Eur. J. Phys. 6~(4) (1985) 287--294.
\newblock \href {http://dx.doi.org/10.1088/0143-0807/6/4/014}
  {\path{doi:10.1088/0143-0807/6/4/014}}.

\bibitem{Abramowitz1964}
M.~Abramowitz, I.~Stegun, {Handbook of Mathematical Functions with Formulas,
  Graphs, and Mathematical Tables}, Dover Publications, Inc., Mineola, NY,
  1964.

\bibitem{Silva2011}
G.~T. Silva, Off-axis scattering of an ultrasound {Bessel} beam by a sphere,
  IEEE Trans. Ultrason. Ferroelect. Freq. Control 58 (2011) 298--304.

\bibitem{Mitri2011}
F.~G. Mitri, G.~T. Silva, Off-axial acoustic scattering of a high-order
  {Bessel} vortex beam by a rigid sphere, Wave Motion 48 (2011) 392--400.

\bibitem{Silva2015}
G.~T. Silva, A.~L. Baggio, J.~H. Lopes, F.~G. Mitri, Computing the acoustic
  radiation force exerted on a sphere using the translational addition theorem,
  IEEE Trans. Ultrason. Ferroelect. Freq. Control 62 (2015) 576--583.

\bibitem{Pao1976}
Y.-H. Pao, V.~Varatharajulu, {Huygens' principle, radiation conditions, and
  integral formulas for the scattering of elastic waves}, J. Acoust. Soc. Am.
  59~(6) (1976) 1361--1371.
\newblock \href {http://dx.doi.org/10.1121/1.381022}
  {\path{doi:10.1121/1.381022}}.

\bibitem{Graff1991}
K.~F. Graff, {Wave Motion in Elastic Solids}, Dover Publications, Inc.,
  Mineola, NY USA, 1991.

\bibitem{Elmore1985}
W.~C. Elmore, M.~A. Heald, {Physics of Waves}, Dover Publications, Inc.,
  Mineola, NY USA, 1985.

\bibitem{Torr1984}
G.~R. Torr, The acoustic radiation force, Am. J. Phys. 52 (1984) 402--408.

\bibitem{Boheren1998}
C.~F. Boheren, D.~R. Huffman, Absorption and Scattering of Light by Small
  Particles, John Wiley \& Sons, Inc., New York NY, USA, 1998.

\bibitem{Westervelt1951}
P.~J. Westervelt, The theory of steady forces caused by sound waves, J. Acoust.
  Soc. Am. 23 (1951) 312--315.

\bibitem{Olsen1958}
H.~Olsen, W.~Romberg, H.~Wergeland, Radiation force on bodies in a sound field,
  J. Acoust. Soc. Am. 30~(1) (1958) 69--76.

\bibitem{Zhang2011}
L.~Zhang, P.~L. Marston, {Geometrical interpretation of negative radiation
  forces of acoustical Bessel beams on spheres}, Phys. Rev. E 84~(3 Pt 2)
  (2011) 035601.
\newblock \href {http://dx.doi.org/10.1103/PhysRevE.84.035601}
  {\path{doi:10.1103/PhysRevE.84.035601}}.

\bibitem{Hulst1981}
H.~C. van~de Hulst, Light Scattering by Small Particles, Dover Publications,
  Inc., New York, NY, 1981.

\bibitem{Brill1987}
D.~Brill, G.~Gaunaurd, Resonance theory of elastic waves ultrasonically
  scattered from an elastic sphere, J. Acoust. Soc. Am. 81 (1987) 1--21.

\bibitem{Ying1956}
C.~F. Ying, R.~Truell, Scattering of a plane longitudinal wave by a spherical
  obstacle in an isotropically elastic solid, J. Appl. Phys. 27~(9) (1956)
  1086--1097.
\newblock \href {http://dx.doi.org/10.1063/1.1722545}
  {\path{doi:10.1063/1.1722545}}.

\bibitem{Einspruch1960}
N.~G. Einspruch, E.~J. Witterholt, R.~Truell, Scattering of a plane transverse
  wave by a spherical obstacle in an elastic medium, J. Appl. Phys. 31~(5)
  (1960) 806--818.
\newblock \href {http://dx.doi.org/10.1063/1.1735701}
  {\path{doi:10.1063/1.1735701}}.

\bibitem{Lucas1982}
B.~G. Lucas, T.~G. Muir, The field of a focusing source, J. Acoust. Soc. Am. 72
  (1982) 1289--1296.

\bibitem{Flax1980}
L.~Flax, H.~\"Uberall, Resonant scattering of elastic waves from spherical
  solid inclusions, J. Acoust. Soc. Am. 67 (1980) 1432--1442.

\bibitem{Newton1982}
R.~G. Newton, Scattering theory of waves and particles, Springer-Verlag, New
  York, USA, 1982.

\bibitem{Kino1987}
G.~S. Kino, Acoustic Waves, Prentice Hall, Englewood Cliffs, NJ, 1987.

\bibitem{Chen2004}
S.~Chen, M.~Fatemi, J.~F. Greenleaf, Quantifying elasticity and viscosity from
  measurement of shear wave speed dispersion, J. Acoust. Soc. Am. 115 (2004)
  2781--2785.

\bibitem{Yamakoshi1990}
Y.~Yamakoshi, J.~Sato, T.~Sato, Ultrasonic imaging of internal vibration of
  soft tissue under forced vibration, IEEE Trans. Ultrason. Ferroelect. Freq.
  Control 37 (1990) 45--53.

\bibitem{Clapp1970}
R.~E. Clapp, Six integral theorems for vector spherical harmonics, J. Math.
  Phys. 11 (1970) 4--9.
\newblock \href {http://dx.doi.org/10.1063/1.1665069}
  {\path{doi:10.1063/1.1665069}}.

\end{thebibliography}

\end{document}